\documentclass[10pt,letterpaper,onecolumn]{article}
\usepackage{amssymb,fullpage,xspace,graphicx,verbatim,amsmath,amsfonts,mathtools,cite,hyperref,tabularx,pbox}
\usepackage[squaren]{SIunits}
\usepackage{titling}

\hypersetup{
    colorlinks=true, % false: boxed links; true: colored links
    linkcolor=red,   % color of internal links
    citecolor=blue,  % color of links to bibliography
    filecolor=blue,  % color of file links
    urlcolor=blue    % color of external links
}

\newcommand{\me}{\text{e}}
\newcommand{\dift}[1]{\frac{\mathrm{d}#1}{\mathrm{d}t}}
\newcommand{\Vol}{S}
\newcommand{\pearson}{Pearson et al.\ \cite{pearson11}\xspace}
\newcommand{\yan}{Yan et al.\ \cite{yan16}\xspace}
\newcommand{\czuppon}{Czuppon et al.\ \cite{czuppon21}\xspace}

\newcommand{\IU}{IV\xspace}
\newcommand{\pevents}{\mathcal{P}_{\text{events}}}
\newcommand{\qfail}{\mathcal{P}_{V \to \, \varnothing}}
\newcommand{\qsuccess}{\mathcal{P}_{V \to \, I }}
\newcommand{\qprod}{\mathcal{P}_{I \to\, m \, V}}
\newcommand{\fsize}{T(\infty)/N_\text{cells}}

\newcommand{\rhoV}{\mathcal{P}_{V \to \, \text{Extinction}}}
\newcommand{\rhoI}{\mathcal{P}_{I \to \, \text{Extinction}}}
\newcommand{\mrhoV}{\mathcal{P}_{V \to \, \text{Establishment}}}
\newcommand{\mrhoI}{\mathcal{P}_{I \to \, \text{Establishment}}}
\newcommand{\Tstar}{T^{*}}
\newcommand{\B}{\mathcal{B}}

\usepackage{lineno}
\usepackage{caption}
\usepackage{enumitem}
{\begin{description}[itemindent=0.5cm]}
{\end{description}}

\newcommand{\mainmstitle}{Stochastic failure of cell infection post viral entry: Implications for infection outcomes and antiviral therapy}

\begin{document}

\title{\mainmstitle}

\author{Christian Quirouette$^1$, Daniel Cresta$^1$, Jizhou Li$^2$, Kathleen P.\ Wilkie$^3$,\\
Haozhao Liang$^{4,\text{\textcurrency}}$, Catherine A.A.\ Beauchemin$^{1,2,*}$\\%
\normalsize
$^1$ Department of Physics, Ryerson University, Toronto, Canada\\
\normalsize
$^2$ Interdisciplinary Theoretical and Mathematical Sciences (iTHEMS), RIKEN, Wako, Japan \\
\normalsize
$^3$ Department of Mathematics, Ryerson University, Toronto, Canada\\
\normalsize
$^4$ Nishina Center for Accelerator-Based Science (RNC), RIKEN, Wako, Japan \\
\normalsize
\textcurrency Current address: Department of Physics, University of Tokyo, Tokyo, Japan\\
\normalsize
$^*$ Corresponding author: \href{mailto:cbeau@ryerson.ca}{cbeau@ryerson.ca}
}

\date{\today}

\maketitle

\begin{abstract}
A virus infection can be initiated with very few or even a single infectious virion, and as such can become extinct, i.e.\ stochastically fail to take hold or spread significantly. There are many ways that a fully competent infectious virion, having successfully entered a cell, can fail to cause a productive infection, i.e.\ one that yields infectious virus progeny. Though many discrete, stochastic mathematical models (DSMs) have been developed and used to estimate a virus infection's extinction probability, these typically neglect infection failure post viral entry. The DSM presented herein introduces parameter $\gamma\in(0,1]$ which corresponds to the probability that a virion's entry into a cell will result in a productive cell infection. We derive an expression for the likelihood of infection extinction in this new DSM, and find that prophylactic therapy with an antiviral acting to reduce $\gamma$ is best at increasing an infection's extinction probability, compared to antivirals acting on the rates of virus production or virus entry into cells. Using the DSM, we investigate the difference in the fraction of cells consumed by so-called extinct versus established virus infections, and find that this distinction becomes biologically meaningless as the probability of extinction approaches 100\%. We show that infections wherein virus is release by an infected cell as a single burst, rather than at a constant rate over the cell's infectious lifespan, has the same probability of infection extinction, despite previous claims to this effect \cite{pearson11}. Instead, extending previous work by others \cite{yan16}, we show how the assumed distribution for the stochastic virus burst size, affects the extinction probability and associated critical antiviral efficacy.
\end{abstract}

%%%%%%%%%%%%%%%%%%%%%%%%%%%%%%%%%%%%%%%%%%%%%%%%%%%%%%%%%%%%%%%%%%%%%%%%%%%%%%%%
%%%%%%%%%%%%%%%%%%%%%%%%%%%%%%%%%%%%%%%%%%%%%%%%%%%%%%%%%%%%%%%%%%%%%%%%%%%%%%%%
%%%%%%%%%%%%%%%%%%%%%%%%%%%%%%%%%%%%%%%%%%%%%%%%%%%%%%%%%%%%%%%%%%%%%%%%%%%%%%%%
%%%%%%%%%%%%%%%%%%%%%%%%%%%%%%%%%%%%%%%%%%%%%%%%%%%%%%%%%%%%%%%%%%%%%%%%%%%%%%%%

\clearpage
\section{Introduction}

Typically, mathematical models describing the course of a virus infection express the number of cells and infectious virions (virus particles) as real positive numbers (continuous), and infection events as deterministic, e.g.\ one virion infects 0.2 cell. By nature, however, the number of cells and virions are whole numbers (discrete) and infection events are stochastic, e.g.\ one virion will infect one cell 20\% of the time. When dealing with large numbers of infectious virions and cells, stochastic fluctuations can become negligible, and the continuous, deterministic, mean-field approach can provide an accurate representation of the kinetics of interest. Yet the small number regime arises commonly. For example, antiviral therapy can reduce the effective number of infection-capable virions to near or below unity. In such cases, random fluctuations could have an important effect on the time course and outcome of an infection.

Several discrete, stochastic mathematical models (DSMs) of viral infection kinetics have been proposed to date, e.g.\ \cite{conway11,pearson11,conway13,heldt15,yan16,conway18,sazonov20}. Notably, Heldt et al.\ \cite{heldt15} predicted with an intracellular DSM that most cells inoculated with a single infectious virion would lead to a non-productive cell infection. But extracellular DSMs typically do not account for cell infection failure post viral entry. When it has been included only some lethal replication errors were represented, e.g.\ lethal reverse transcription errors only \cite{conway18}. In fact, there are many ways that infectious virions post cell entry can fail to cause a cell infection that will produce fully infectious progeny.

For human immunodeficiency virus (HIV), entry begins with successful binding to the cell surface, followed by fusion of the virus membrane with the cell surface. Yet sometimes virions can be taken in via the endocytic route \cite{permanyer10}. In this case, virions can fail to fuse with late endosomes and eventually be degraded \cite{fredericksen02}. Or, if there is successful entry, virions can still be degraded in the cytosol \cite{wei05}. The virus can also fail to be imported into the nucleus \cite{burdick13,burdick17}. In addition, fatal mutations can be acquired during reverse transcription \cite{ho13}, when viral RNA is transcribed into complementary DNA (cDNA) which is used to make typically a single copy of viral DNA for integration. Finally, failure may also be the result of host cell mechanisms that interfere with early replication steps \cite{colomerlulch18,kieffer05}.

For influenza A virus (IAV), entry begins with successful binding to the cell surface, followed by internalization by endocytosis. Failure to undergo fusion with late endosomes can then occur \cite{koff79,qin19}. If uncoating happens, released viral ribonucleoproteins (vRNPs) containing the viral RNA segments could be subject to RNA degradation in the cytosol \cite{schelker16} or fail to enter the nucleus \cite{flatt15,qin19}. Degradation of one or more genome segments, following nuclear import and prior to transcription could also occur \cite{heldt15}. There is also a multitude of host cell mechanisms that can impair early steps of viral replication \cite{xiao13mxa,desai14,villalon17} and these may be yet more potential sources for failure. 

These are just some of the ways in which otherwise fully physically infectious virions could, through random chance, fail to complete a key step following cell entry. In addition to such stochastic occurrences in fully functional virions, a number of entry capable virions could have physical defects that prevents them from completing one or more key replication steps, leaving them physically unable to cause a productive infection. As such, failure of productive cell infection post viral entry is a combination of both the stochastic post-entry failure of otherwise fully infectious virions, and the inevitable failure of replication defective, entry capable virions.

A common, important application of DSMs is to estimate the extinction probability of an infection: the likelihood that the infection will fail to take hold or spread significantly. It typically depends on both the replication capabilities of the virus (i.e.\ infection parameters) and the initial number of infectious virions or cells. The extinction probability is an important quantity to derive as it can, for example, be used to evaluate the probability of success of antiviral therapy \cite{conway13,czuppon21}. In particular, \czuppon compared the ability of prophylactic antivirals acting either on the rate of virus entry into cell or the rate of virus production, to reduce the establishment probability, or $1-$(extinction probability), of severe acute respiratory syndrome coronavirus 2 (SARS-CoV-2) infection in vivo. However, the effect of certain classes of antivirals, such as endosomal fusion inhibitors, would be better represented as reducing the probability that, having successfully entered a cell, a virion will go on to cause the productive infection of that cell. For influenza A virus, it would be a more appropriate way to represent the mode of action of adamantanes, such as amantadine and rimantadine which block the M2 ion channel necessary for successful fusion of influenza A virus with the endosome \cite{dobrovolny17,cbeau08}. For SARS-CoV-2, it has been suggested that cathepsin L inhibitors can reduce the probability of successful endosomal fusion \cite{liu20}. Failure to undergo endosomal fusion would lead to loss of the infectious virion, but would not result in a productive cell infection.

Although the extinction probability is often an important consideration in comparing prophylactic antivirals, what is never discussed is the number or fraction of cells that are actually consumed by infections that are said to have gone ``extinct'' or to have ``established''. It is possible that antivirals with different modes of action, even with the same extinction probability, could result in a very different fraction of cells consumed by so-called established or extinct infections. This would have important implications for how one should interpret the probability of success of a particular antiviral therapy.

In addition, nearly all DSMs, including past works that used the extinction probability to evaluate the probability of success of antiviral therapy \cite{conway13,czuppon21}, assume the duration of the infectious phase, the period during which infected cells are producing and releasing virus progeny, to be exponentially-distributed. Careful pairing of mathematical models and experimental measurements has established that the duration of the infectious phase in vitro for cells infected with IAV \cite{holder11}, simian HIV \cite{cbeau17}, or Ebola virus \cite{liao20}, follows a log-normal or normal-like distribution, and has clearly rejected the probability of an exponentially distributed infectious phase duration. \yan estimated the extinction probability for a DSM of IAV infection in vivo that allowed the lifespan of infectious cells to follow an Erlang distribution which, via its shape parameter ($k$), can capture exponential ($k=1$), log-normal-like ($k\sim[1,6]$), normal-like ($k>10$), and even Dirac delta-like ($k\to\infty$) distributions. \yan have shown that increasing the shape parameter ($k$) from exponential to log-normal leads to a decrease in the extinction probability given an infection initiated with a number of infectious virions. Hence, the distribution of the infectious phase duration is also expected to affect the likelihood that an infection will become established under antiviral therapy.

In this work, we construct a DSM for virus infection through the physical consideration of key infection steps, rather than through a systematic, direct mathematical conversion of our mean-field model. Our DSM explicitly represents the probability that a virion, after having successfully entered a cell, will fail to result in the productive infection of that cell. The DSM is first used to estimate the extinction probability of an infection. Extending work by \czuppon, we show that prophylactic therapy with an antiviral that blocks productive cell infection \emph{after} viral entry, is better at reducing the establishment probability, than one acting to reduce either the virus production rate or the rate of virus entry into cells. In addition, we investigate the difference in the fraction of cells consumed by so-called extinct versus established infections, and re-visit the comparison of antivirals though this new lens. Finally, we demonstrate how the antiviral efficacy required to achieve a desired infection extinction probability critically depends on the assumed distribution of the infectious phase duration.

%%%%%%%%%%%%%%%%%%%%%%%%%%%%%%%%%%%%%%%%%%%%%%%%%%%%%%%%%%%%%%%%%%%%%%%%%%%%%%%%
%%%%%%%%%%%%%%%%%%%%%%%%%%%%%%%%%%%%%%%%%%%%%%%%%%%%%%%%%%%%%%%%%%%%%%%%%%%%%%%%
%%%%%%%%%%%%%%%%%%%%%%%%%%%%%%%%%%%%%%%%%%%%%%%%%%%%%%%%%%%%%%%%%%%%%%%%%%%%%%%%
%%%%%%%%%%%%%%%%%%%%%%%%%%%%%%%%%%%%%%%%%%%%%%%%%%%%%%%%%%%%%%%%%%%%%%%%%%%%%%%%

\clearpage
\section{Results}

\subsection{The mathematical models}

The mean-field mathematical model (MFM) and its DSM counterpart used herein are given by
\begin{align} \label{mfeq}
\dift{T} &= -\gamma \beta_\text{} TV_\text{}/\Vol  &T^{t+1} &= T^t-N^{\text{inf}} \nonumber \\
\dift{E_1} &= \gamma \beta_\text{} TV_\text{}/\Vol - \frac{n_E}{\tau_E} E_1 &E_1^{t+1} &= E_1^t + N^{\text{inf}} - E_1^{\text{out}} \nonumber \\
\dift{E_i} &= \frac{n_E}{\tau_E}E_{i-1} - \frac{n_E}{\tau_E}E_i &E_i^{t+1} &= E_i^t + E_{i-1}^{\text{out}} - E_i^{\text{out}} \qquad i=2,3,...,n_E \\
\dift{I_1} &= \frac{n_E}{\tau_E}E_{n_E}-\frac{n_I}{\tau_I}I_1 &I_1^{t+1} &= I_1^t + E_{n_E}^{\text{out}} - I_1^{\text{out}} \nonumber \\
\dift{I_j} &= \frac{n_I}{\tau_I}I_{j-1}-\frac{n_I}{\tau_I}I_j &I_j^{t+1} &= I_j^t + I_{j-1}^{\text{out}} - I_j^{\text{out}} \qquad j=2,3,...,n_I \nonumber \\
\dift{V} &= p_\text{}\sum_{j=1}^{n_I}I_j-cV_\text{}-\beta_{} TV_\text{}/\Vol &V^{t+1} &= V^\text{prod} + V^\text{remain} \nonumber
\end{align}

Initially, all cells are uninfected, susceptible target cells, $T$, i.e.\ $T(t=0) = N_\text{cells}$. Target cells, $T$, are then infected by infectious virions, $V$, resulting in a successful cell infection ($T\rightarrow E_1$) at rate $\gamma\beta V/\Vol$, where $\Vol$ is the volume of supernatant. Newly infected cells enter ($T\rightarrow E_1$) and traverse ($E_1\rightarrow E_2\rightarrow\ldots\rightarrow E_{n_E}$) the $n_E$ compartments of the eclipse phase, during which cells are infected but are not yet producing infectious virions. Infected cells then enter ($E_{n_E}\rightarrow I_1$) and traverse ($I_1\rightarrow I_2\rightarrow\ldots\rightarrow I_{n_I}$) the infectious phase, during which they produce infectious virions at constant rate $p$. As infected cells leave the last compartment ($I_{n_I}$), they cease virus production and thus cease to contribute to the infection kinetics, and possibly undergo apoptosis. The exponentially-distributed durations of the $n_E$ eclipse (or $n_I$ infectious) phase compartments together yield an Erlang-distributed total duration for the eclipse (or infectious) phase of mean duration $\tau_E$ (or $\tau_I$), and standard deviation $\tau_E/\sqrt{n_E}$ (or $\tau_I/\sqrt{n_I}$), where $n_E$ (or $n_I$) corresponds to the shape parameter of the Erlang distribution. Infectious virions ($V$) are produced at a constant rate of $p$ per infectious cell per hour, and are lost either through loss of infectivity at rate $c$, or entry into susceptible cells at rate $\beta T/\Vol$.

This MFM is similar to that widely validated and applied to analyze and predict the course of in vitro infections with IAV \cite{paradis15,holder12compete,yan20}, simian HIV \cite{cbeau17}, Ebola virus \cite{liao20}, RSV \cite{gonzalez18compare} and rotavirus \cite{gonzalez18}. It differs from the latter by explicitly accounting for the loss of infectious virions due to cell entry at rate $\beta TV/\Vol$, which causes a corresponding loss of uninfected target cells becoming infected ($T\rightarrow E_1$) at rate $\gamma\beta TV/\Vol$. Parameter $\gamma$ therefore has units of cell per infectious virion (\IU), where $\gamma\in\unit{(0,1]}{cell/\IU}$, and corresponds to the average fraction of infectious virion entries into a cell that results in a successful infection of the cell ($T\rightarrow E_1$). This parameter is not typically included in ordinary differential equation (ODE) models of viral infection kinetics.

Biologically, $\gamma$ accounts for several different causes of infection failure post viral entry into a cell. It can represent semi-infectious virions that are entry-competent, but are defective in their ability to complete another downstream step, e.g.\ are missing one or more viral genome segment or have deleterious genetic mutations. It can also represent fully infectious virions that are not defective but rather, through random chance, fail to achieve a key step they could have functionally achieved, e.g.\ 50\% of influenza A virions failing to fuse with endosome membrane following cell entry \cite{koff79,stegmann93}. Herein it is assumed that when an infectious virion enters a cell, the cell will either be successfully infected ($T\rightarrow E_1$) with probability $\gamma$, or remain uninfected ($T$) with probability $1-\gamma$. Cells being left in a semi-infected state by partially failed infections are not considered \cite{liao16,brooke13,diefenbacher18}.

The DSM is largely analogous to the MFM, where DSM variables, e.g.\ $T^t$ the number of target cells at time $t$, each denoted with a superscript $t$, are whole numbers. The remaining terms, corresponding to changes in these variables, are random whole numbers drawn at each time step from distributions thought to best represent the corresponding underlying biological process where

\begin{itemize}
\item $E_i^\text{out} = \text{Binomial}(n=E_i^t,\, p_E= \Delta t\cdot n_E/\tau_E)$ corresponds to the number of cells in the $i^\text{th}$ eclipse compartment ($E_i$) that will transition to the $(i+1)^\text{th}$ compartment over a time interval $\Delta t$, given that there are $E_i^t$ cells in the $i^\text{th}$ compartment at time step $t$. Each of the $E_i^t$ cells in eclipse compartment $i$ undergo an independent Bernoulli trial with two possible outcomes: either the cell transitions to the $(i+1)^\text{th}$ compartment with success probability $p_E=\Delta t\cdot n_E/\tau_E$, or otherwise remains in the $i^\text{th}$ compartment. The time step, $\Delta t$ is chosen to be sufficiently small to ensure $p_E < 1$, as discussed below.

\item $I_j^\text{out} = \text{Binomial}(n=I_j^t,\, p_I= \Delta t\cdot n_I/\tau_I)$ is the number of cells in the $j^\text{th}$ infectious compartment ($I_j$) that transition to the $(j+1)^\text{th}$ compartment over a time interval $\Delta t$, given that there are $I_j^t$ cells in the $j^\text{th}$ compartment at time step $t$, and the probability of transition $p_I= \Delta t\cdot n_I/\tau_I$.

\item $V^\text{prod} = \text{Poisson}(\lambda=\Delta t\cdot p\,\sum_{j=1}^{n_I} I_j^t)$ is the number of infectious virions newly produced into the supernatant over a time $\Delta t$, given that there are $\sum_{j=1}^{n_I} I_j^t$ infectious cells at time step $t$, each producing infectious virions at a rate of $p$ infectious virions per hour. The Poisson-distributed random variable, $V^\text{prod}$ is the number of events that occurred, given the expected number of occurrences over a time $\Delta t$, namely $\lambda=\Delta t\cdot p\sum_{j=1}^{n_I} I_j^t$.

\item $V^\text{decay},V^\text{enter},V^\text{remain} = \text{Trinomial}(n=V^t,\, p_1=\Delta t\cdot c,\, p_2=\Delta t\cdot\beta T^t/\Vol,\, p_3=1-p_1-p_2)$ corresponds to the number of infectious virions in the supernatant at time $t$, $V^t$, that end up in each of 3 possible fates, namely $V^t = V^\text{decay}+V^\text{enter}+V^\text{remain}$. $V^\text{decay}$ is the number of infectious virions that lose infectivity with probability $p_1=\Delta t\cdot c$, $V^\text{enter}$ are lost from the supernatant as they enter a target cell with probability $p_2=\Delta t\cdot\beta T^t/\Vol$, and $V^\text{remain}$ infectious virions do neither and remain in the supernatant with probability $p_3=1-p_1-p_2$. Each probability ($p_1,p_2,p_3$) can be considered constant over time interval $\Delta t$, provided a sufficiently small time step is chosen, as discussed below.

\item $N^\text{inf} = \texttt{len(numpy.unique(numpy.random.choice(a=}T^t\texttt{,size=}V^\text{suc}\texttt{,replace=True)))}$ is the number of target cells that become infected ($T\rightarrow E_1$) over time $\Delta t$, given that $V^\text{enter}$ infectious virions enter into target cells, out of which $V^\text{suc} = \text{Binomial}(n=V^\text{enter},\, p_V=\gamma)$ infectious virions ultimately lead to successful cell infection, given probability $p_V=\gamma$. Random variable $N^\text{inf}$ does not correspond to any named probability mass function. The \texttt{Python} expression simulates randomly placing $V^{\text{suc}}$ infectious virions into $T^t$ cells chosen at random with replacement, \texttt{numpy.random.choice}, and counts the number of distinct cells that received one or more infectious virions, \texttt{len(numpy.unique(...))}.
\end{itemize}

\begin{table*}
\begin{center}
\caption{Random variables of the DSM}
\label{summary-rvar}
\begin{tabular}{ll}
Random variable & Random number generator \\
\hline
$E_i^\text{out}$ & $\text{Binomial}(n=E_i^t,\, p_E=\Delta t\cdot n_E/\tau_E)$ where $i=1,2,...,n_E$\\
$I_j^\text{out}$ & $\text{Binomial}(n=I_j^t,\, p_I=\Delta t\cdot n_I/\tau_I)$ where $j=1,2,...,n_I$\\
$V^\text{prod}$ & $\text{Poisson}(\lambda=\Delta t\cdot p \sum_{j=1}^{n_I} I_j^t)$ \\
$V^\text{decay},V^\text{enter},V^\text{remain}$ & $\text{Trinomial}(n=V^t,\, p_1=\Delta t\cdot c,\, p_2=\Delta t\cdot\beta T^t/\Vol,\, p_3=1-p_1-p_2)$ \\
$N^\text{inf}$ & $\texttt{len(numpy.unique(numpy.random.choice(a=}T^t\texttt{,size=}V^\text{suc}\texttt{,replace=True)))}$ \\
& \hspace{2em} where $V^\text{suc} = \text{Binomial}(n=V^{\text{enter}},\,p_V=\gamma)$ \\
\hline
\end{tabular}
\end{center}
\end{table*}

Table \ref{summary-rvar} summarizes how each random variable is generated. The duration of the DSM's discrete time steps, $\Delta t$, which sets the probability of event occurrences or the number of such events, is computed at each iteration step $t$ as
\begin{align}
\Delta t = \frac{\pevents}{
\max \bigg\{ \frac{\beta T^t}{\Vol}, c, \frac{n_E}{\tau_E}, \frac{n_I}{\tau_I} \bigg\} }
\end{align}
where $\pevents$ is the probability of occurrence of the most likely event, namely of virion loss due to cell entry ($\beta T^t/\Vol$) or loss of infectivity ($c$), or transition of cells from one infected state to another ($n_E/\tau_E$ and $n_I/\tau_I$). A value of $\pevents=0.05$ (or 5\%) is used as it was found to be small enough that choosing a smaller value did not affect the results presented herein. The numerical solution of the DSM was further validated by replacing the DSM's random variables, e.g.\ $E_i^\text{out}$, by their expected value, e.g.\ $E_i^\text{out} = E_i^t\cdot\Delta t\cdot n_E/\tau_E$, (see Methods, Section \ref{mf-solver-expected-val}) and comparing the solution against that obtained for the MFM in Eqn.\ \eqref{mfeq} with a standard numerical ODE solver. The solutions were found to be in agreement over the wide range of parameter values explored (not shown).

Some readers might prefer to see the DSM expressed as transitions \cite{pearson11,yan16}, namely
\begin{align}
V &\xrightharpoonup[]{\mathmakebox[1cm]{\beta T/S}}  V^\text{enter} \nonumber \\
V^\text{enter}+T &\xrightharpoonup[]{\mathmakebox[1cm]\gamma} E_1 \nonumber \\
V^\text{enter}+T &\xrightharpoonup[]{\mathmakebox[1cm]{1-\gamma}} T \nonumber \\
E_i &\xrightharpoonup[]{\mathmakebox[1cm]{n_E/\tau_E}} E_{i+1} \qquad i=1,2,3,...,n_E-1 \nonumber \\
E_{n_E} &\xrightharpoonup[]{\mathmakebox[1cm]{n_E/\tau_E}} I_1 \\
I_j &\xrightharpoonup[]{\mathmakebox[1cm]{n_I/\tau_I}} I_{j+1} \qquad j=1,2,3,...,n_I-1 \nonumber \\
I_{n_I} &\xrightharpoonup[]{\mathmakebox[1cm]{n_I/\tau_I}} \varnothing \nonumber \\
I_j &\xrightharpoonup[]{\mathmakebox[1cm]p} I_j+V \qquad j=1,2,3,...,n_I \nonumber \\
V &\xrightharpoonup[]{\mathmakebox[1cm]c}  \varnothing \nonumber 
\end{align}
One limitation of this representation is that it indicates that one virion that enters a cell and that is successful at causing an infection with rate $\gamma$ will cause the infection of one cell ($V^\text{enter}+T \xrightharpoonup[]{} E_1$). In reality, in our DSM, one successful virion does not necessarily cause the infection of one cell (see $N^\text{inf}$ in Table \ref{summary-rvar}) because one cell could receive two successful infectious virions which would only result in one rather than two new successful cell infections.

\subsection{Important biological quantities}

Following \pearson, let us define $\rhoV$ and $\rhoI$ as the probability of infection extinction given an infection initiated with either only one infectious virion or only one infectious cell, respectively. The extinction probability given any initial number $V_0$ of infectious virions and $I_0$ of infectious cells is then $\left(\rhoV\right)^{V_0} \cdot \left(\rhoI\right)^{I_0}$.

If initially there is only one infectious virion, the infection can fail to spread if either that initial infectious virion fails to cause a successful cell infection with probability $\qfail$, or if it does cause a successful cell infection with probability $\qsuccess=(1-\qfail)$ but that cell infection subsequently leads to extinction with probability $\rhoI$. We can therefore write
\begin{align} \label{rhoVa}
\rhoV = \qfail+\qsuccess \cdot \rhoI \ .
\end{align}

If initially there is only one infectious cell, the cell will produce $m$ infectious virions over its lifespan with probability $\qprod$. Each one of these produced infectious virions can be treated as an independent infection event such that,
\begin{align}
\rhoI = \sum_{m=0}^\infty \qprod \cdot \left(\rhoV\right)^m \ .
\end{align}

Taken together, the extinction probability given an infection initiated with only one infectious virion, $\rhoV$, yields the recursive relation
\begin{align} \label{rhoVr}
\rhoV = \qfail+\qsuccess \cdot \sum_{m=0}^\infty \qprod \cdot \left(\rhoV\right)^m \ .
\end{align}

The probability that an infectious virion is successful at causing a productive cell infection, $\qsuccess$, can be expressed as the ratio between the rate of successful cell infection per infectious virion $\gamma\beta N_\text{cells}/\Vol$ and the rate of virion loss through loss of infectivity plus cell entry, $c+\beta N_\text{cells}/\Vol$ (see Methods, Section \ref{qsuccess-section} for derivation), namely,
\begin{align} \label{qsuccr}
\qsuccess = (1-\qfail) = \frac{\gamma\beta N_\text{cells}/\Vol}{c+\beta N_\text{cells}/\Vol} \ .
\end{align}

The probability that an infectious cell produces $m$ infectious virions over its lifespan, $\qprod$, can be expressed as the marginal probability distribution of the probability that an infectious cell produces $m$ infectious virions given a lifespan of length $t$, $\text{Poisson}(m|\lambda=pt)$, and the probability that the lifespan is of length $t$, $\text{Erlang}(t|k=n_I,\lambda=n_I/\tau_I)$. Therefore, 
\begin{align}
\qprod
&= \int_0^\infty \text{Poisson}(m|\lambda=pt) \cdot \text{Erlang}(t|k=n_I,\, \lambda=n_I/\tau_I)\ \text{d}t \nonumber \\
&= \frac{(m+r-1)!}{m!(r-1)!} (1-p_B)^r (p_B)^m
= \text{NB}(m|r=n_I,\, p_B=\B/(n_I+\B)) \label{burstr}
\end{align}
where NB stands for the negative binomial (or Pascal) distribution and describes the probability that $m$ successes will have occurred by the time one has observed $r=n_I$ failures from a series of independent Bernoulli trials with a probability of success $p_B$ (see Methods, Section \ref{qprod-section} for derivation). The mean of the distribution corresponds to the average burst size, $\B = p \tau_I$, i.e.\ the average number of infectious virions produced by a productively infected cell over its lifespan, with $p$ the rate of infectious virion production per cell and $\tau_I$ the average infectious cell lifespan.

Combining the mean number of infectious virions produced by a productively infected cell, $\B$, and the probability that each of them causes the infection of a cell, $\qsuccess$ (Eqn.\ \eqref{qsuccr}), we obtain the average basic reproductive number ($R_0$), defined as the average number of productive secondary infections caused by an infected cell over its lifespan when placed within a fully susceptible and uninfected cell population,
\begin{align}
R_0 = \B \cdot \qsuccess
= p \tau_I \cdot \frac{\gamma\beta N_\text{cells}/\Vol}{c+\beta N_\text{cells}/\Vol} \label{R0} \ .
\end{align}
Substituting Eqn.\ \eqref{qsuccr} and Eqn.\ \eqref{burstr} into Eqn.\ \eqref{rhoVr} results in the following expression for $\rhoV$
\begin{align}
0 &= 1-\frac{1+c/(\beta N_\text{cells}/\Vol)}{\gamma}[1-\rhoV]-\left[\frac{\B(1-\rhoV)}{n_I}+1\right]^{-n_I} \label{rootm} \ .
\end{align}
While Eqn.\ \eqref{rootm} does not lead to an analytical solution for $\rhoV$, a solution can easily be obtained numerically (see Methods, Section \ref{rhoV-section}). The probability of infection establishment given an infection initiated with only one infectious virion, $\mrhoV$, is simply $1-\rhoV$.

\subsection{Effectiveness of antivirals to reduce an infection's establishment probability}

Prophylactic antiviral therapies are often characterized and compared in terms of their ability to reduce an infection's establishment probability. This is because for natural infections in a host, the initial virus inoculum can be sufficiently small that prophylactic antiviral therapy can prevent infection, i.e.\ induce infection extinction.

Recently, \czuppon used a DSM to evaluate the effectiveness of prophylactic treatment with antivirals acting on the rate of virus entry into cells sometimes called the virus infectivity rate, herein $\beta$, or the virus production rate, $p$, to reduce the establishment probability for a SARS-CoV-2 infection in vivo initiated by a small number of infectious virions. But an endosomal fusion inhibitor (e.g., cathepsin L inhibitors for SARS-CoV-2 \cite{liu20}) would lead to the removal of an infectious virion from the medium (i.e., leave $\beta$ unaffected), and yet by causing fusion failure post cell entry, would block the cell from progressing to a productively infectious state, hence blocking 100\% of viral progeny production. Using our DSM, which can explicitly represent productive infection failure post viral entry via parameter $\gamma$, we extend Czuppon's investigation to evaluate how such an antiviral would compare against those acting on $\beta$ or $p$.

\begin{figure} 
\begin{center}
\includegraphics[width=0.55\linewidth]{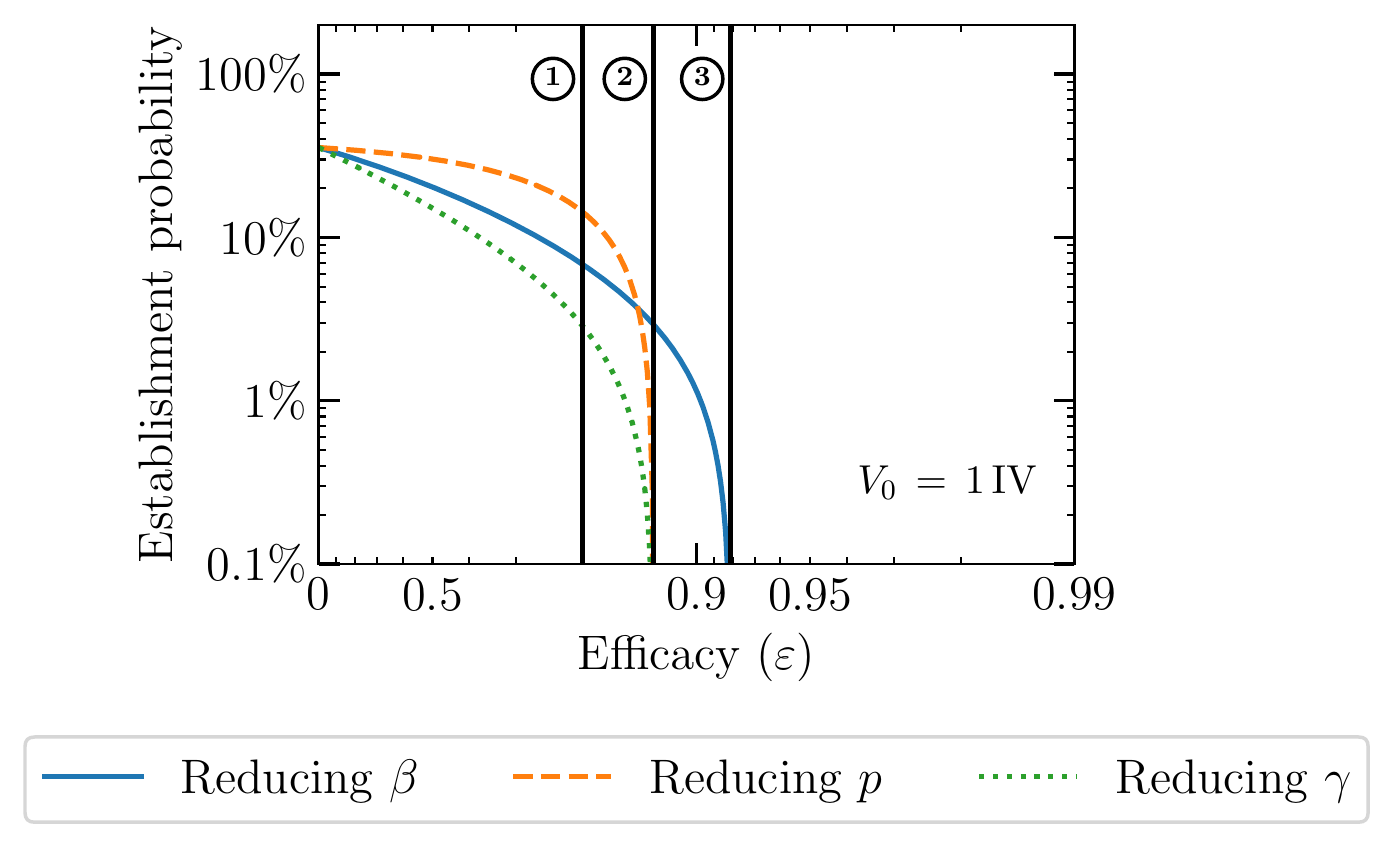}
\caption{\textbf{Evaluating the ability of antivirals to reduce an infection's establishment probability.} The establishment probability given an infection initiated with only one infectious virion as a function of efficacy ($\varepsilon$) for antivirals acting either to reduce the virus entry rate, $\beta \to (1-\varepsilon)\beta$ (blue solid line), the virus production rate, $p \to (1-\varepsilon)p$ (orange dashed line), or the probability of a successful cell infection post viral entry, $\gamma \to (1-\varepsilon)\gamma$ (green dotted line). The line labelled 1 indicates a value of lower efficacy $\varepsilon = 0.8$. The line labelled 2 indicates a value of higher efficacy where the establishment probability goes to zero for an antiviral acting on $p$ or $\gamma$. The line labelled 3 indicates an even higher value of efficacy needed for the establishment probability to go to zero for an antiviral acting on $\beta$. The infection parameters are from \czuppon (see Methods, Section \ref{summary-params}).}
\label{mrhoV}
\end{center}
\end{figure}

Fig~\ref{mrhoV} shows the establishment probability given an infection initiated with only one infectious virion as a function of antiviral efficacy ($\varepsilon$) for antivirals acting on $\beta$, $p$ or $\gamma$. The parameter values used to generate this and all other figures herein are provided in the Methods, Section \ref{summary-params}. The infection parameters were taken from \czuppon, and in particular assume an exponentially distributed duration for the eclipse and infectious phases ($n_E=n_I=1$). For $n_I=1$, the expression for the establishment probability in Eqn.\ \eqref{rootm} reduces to
\begin{align}
\mrhoV = \qsuccess - \frac{1}{\B} = \frac{R_0}{\B} - \frac{1}{\B} =  \frac{\gamma\beta N_\text{cells}/\Vol}{c+\beta N_\text{cells}/\Vol} - \frac{1}{p \tau_I}
\label{mrhoVexp}
\end{align}
where $\B=p\tau_I$ is the burst size (see Methods, Section \ref{rhoV-section}).

At relatively low efficacy, e.g.\ $\varepsilon = 0.8$ (see line 1 in Fig~\ref{mrhoV}), for the parameters chosen, an antiviral acting on $\gamma$ is more effective than one acting on $\beta$ which is more effective than one acting on $p$ at reducing the establishment probability. The infection parameters used by \czuppon correspond to a sufficiently large burst size ($\B=p\tau_I=18.8$, see Methods Section \ref{summary-params}) that $\mrhoV\approx\qsuccess$ ($=0.409$). As such, the bottleneck to infection establishment is the probability that the single initial infectious virion causes a productive cell infection, $\qsuccess$, which depends on $\beta$ and $\gamma$, but not $p$. While an antiviral acting on $\beta$ reduces both the numerator and denominator of $\qsuccess$, one acting on $\gamma$ reduces only the numerator and therefore has a greater effect. At this relatively low antiviral efficacy ($\varepsilon\approx0.8$), as the initial number of infectious virions increases and as the establishment probability approaches 100\%, differences in the efficacy of antivirals acting on $p$, $\beta$ or $\gamma$ to reduce the establishment probability vanishes (see Supplementary Material \ref{V0-section}).

As the antiviral efficacy is increased further, it eventually reaches the critical point where $R_0=1$, and the establishment probability, $(R_0-1)/\B$, equals zero. The value of the antiviral efficacy at which this is achieved is the same for antivirals acting on $p$ and $\gamma$ (see line 2 in Fig~\ref{mrhoV}) since they affect $R_0$ (Eqn.\ \eqref{R0}) identically. The efficacy of an antiviral acting on $\beta$ must be much higher in order to reach this critical point (see line 3 in Fig~\ref{mrhoV}), because reducing $\beta$ reduces both the numerator and denominator of $R_0$. 

Therefore, as \czuppon before us, and for their choice of parameters, we find that at lower efficacies, an antiviral acting on $\beta$ is better than one acting on $p$ at reducing the establishment probability, whereas at higher efficacies it is the opposite. However, here we show that an antiviral acting on $\gamma$ at any efficacy is best, better than one acting on $p$ or $\beta$, at reducing the establishment probability.

\subsection{Distinction between infection extinction and establishment}

What constitutes the establishment or extinction of an infection is commonly, somewhat vaguely defined. For example, \czuppon state \emph{``The larger the initial inoculum of infectious virus, the less likely is the prevention of an infection.''} equating infection extinction as predicted by their DSM with prevention of the infection. \pearson state \emph{``Whether exposure to virus leads to systemic infection or complete elimination of the virus can be a matter of luck, particularly when exposure is to low levels of virus.''} and later define the extinction probability as \emph{``the probability that the virus and all infected cells are completely eliminated from the host''}. While it is never explicitly stated, one could be forgiven for thinking that extinction means that almost no cells are consumed while establishment means that all but a negligible number of cells are left untouched by the infection. But to what extent is this true? Is the distinction always biologically meaningful and clear? Can one look at an infection outcome and clearly categorize it as an extinct versus an established infection?

\begin{figure}
\begin{center}
\includegraphics[width=1.0\linewidth]{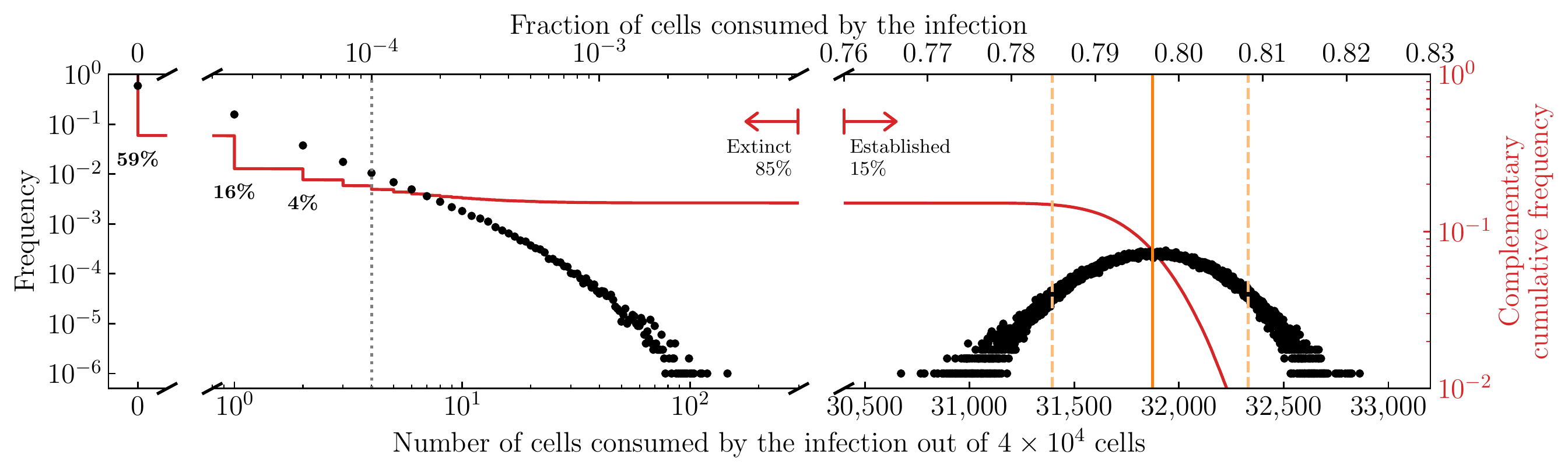}
\caption{\textbf{Infection extinction versus establishment.}
The distribution for the frequency (black dots) or complementary cumulative frequency (solid red curve), or $1-$(cumulative frequency), of the number of cells consumed by the infection, $N_\text{cells}-T(\infty)$, out of $N_\text{cells} = \unit{4 \times 10^4}{cells}$, based on $10^6$ DSM simulations. For the frequency distribution, each black dot represents the fraction of the DSM simulations where exactly this discrete number of cells were consumed by the infection (e.g.\ $0,1,2,...$). The vertical solid (orange) line represents the MFM-predicted fraction of cells consumed by an established infection. The 95\% ($2\sigma$) CI for the number of cells consumed by infections corresponds to the region left of the vertical (grey) dotted line for extinct infections, or enclosed by a pair of vertical (orange) dashed lines for established infections. Parameters were the same as in Fig~\ref{mrhoV} for an antiviral acting on the virus production rate, $p\to(1-\varepsilon)p$, at efficacy $\varepsilon=0.79$ such that $\Tstar/N_\text{cells} = 0.5$ (Eqn.\ \eqref{tstar}).}
\label{example}
\end{center}
\end{figure}

Fig~\ref{example} shows the frequency and complementary cumulative frequency distributions, or $1-$(cumulative frequency), for the number of cells consumed by $10^6$ DSM simulated infections. The DSM simulations appear to fall into 2 distinct categories: infections with relatively few (close to 0\% on the left) or many (around 80\% on the right) cells consumed, naturally corresponding to infections said to have gone extinct or to have established, respectively, as indicated by red arrows. Extinct infections make up 85\% (848,045) of the $10^6$ DSM simulations, in agreement with the theoretical value of the extinction probability, $\rhoV$, for the infection parameters used. Of the extinct infections, 69\% (59\%/85\%) result in no cells consumed, 93\% ($(59+16+4)\%/85\%$) in fewer than 3 cells consumed, and 95\% ($2\sigma$) in fewer than \unit{5}{cells} consumed, or less than 0.01\% of all susceptible cells. With so few cells infected, it could be reasonable to expect that no significant immune response nor symptoms were triggered. Of the established infections, all consumed at least 30,600 cells or about 10,000$\times$ more than infections considered extinct. The difference between extinction and establishment in this example is therefore statistically and biologically significant.

For established infections, the MFM predicts that $\sim$80\% (or 31,873 out of 40,000 cells) of cells will be consumed, given by $1-T(\infty)/N_\text{cells}$ in Eqn.\ \eqref{fsize3}, and indicated by a vertical solid line in Fig~\ref{example}. This roughly corresponds to the median of the distribution of cells consumed by established DSM infections --- also its mean and mode when the distribution is symmetric, as is the case here --- thus providing an analytical expression to track this key feature of the DSM distribution.

\begin{figure}
\begin{center}
\includegraphics[width=1.0\linewidth]{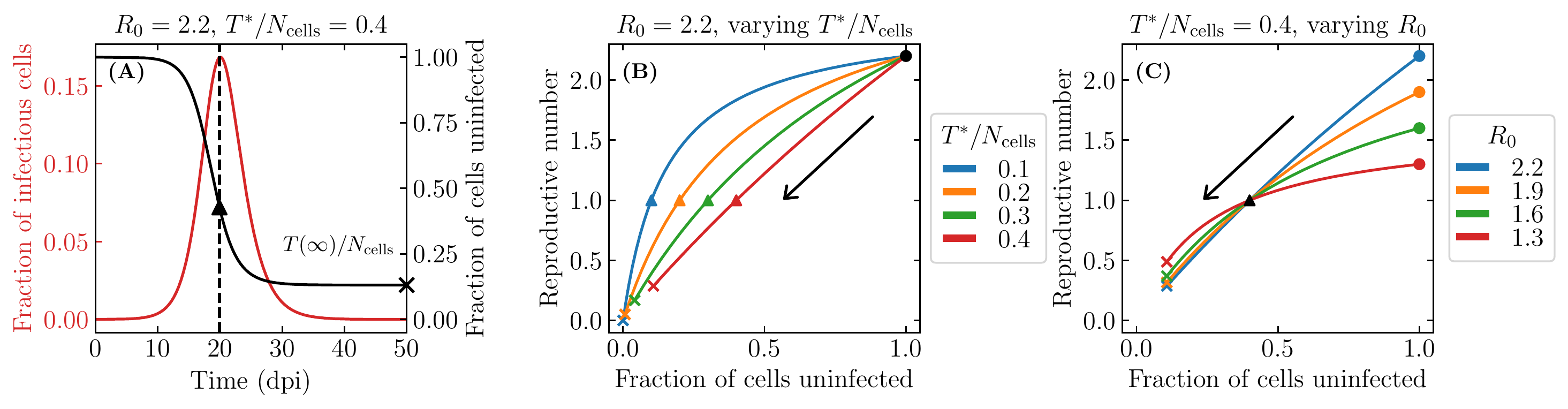}
\caption{\textbf{Exploring the critical fraction of uninfected cells, $\Tstar/N_\text{cells}$.}
(A) MFM-predicted time courses for the fraction of uninfected (black) and infectious (red) cells, for an infection initiated with 10 infectious virions under antiviral therapy acting to reduce the virus entry rate, $\beta \to (1-\varepsilon)\beta$, at efficacy $\varepsilon = 0.81$, which yields an establishment probability ($\mrhoV=48\%$) that is $\sim$50\% of its value without antivirals \cite{czuppon21}. The point $(t,\Tstar/N_\text{cells})$ is represented by a triangle; the time when $T = \Tstar$ by a dashed line; and $T=T(\infty)$ by an $\times$ on the right vertical axis. The parameters are provided in Methods, Section \ref{summary-params}, but notably $n_I=1$.
(B,C) The reproductive number, $R(t)$, as a function of the fraction of cells that remain uninfected, $T(t)$, over the course of an infection (time is implicit), based on Eqn.\ \eqref{phaseeq} when either $R_0 = 2.2$ and $\Tstar/N_\text{cells}$ is varied (B); or $\Tstar/N_\text{cells} = 0.4$ and $R_0$ is varied (C). The start of the infection is represented by a circle; the critical point where $T=\Tstar$ and $R = 1$ by a triangle; and the end of the infection where $T=T(\infty)$ as given by Eqn.\ \eqref{fsize4} by an $\times$.}
\label{phase}
\end{center}
\end{figure}

Let us consider how this MFM-predicted fraction of cells consumed by established infections depends on infection parameters (see Methods, Section \ref{final-size-section} for derivation). At the start of an infection, all cells are uninfected, $T(0) = N_\text{cells}$, and the reproductive number, $R(0)$, i.e.\ the number of successful secondary infections caused by an infected cell over its lifespan, corresponds to the basic reproductive number ($R_0$), given by Eqn.\ \eqref{R0}, where $R_0>1$ for an infection that has a non-zero establishment probability. As the infection progresses and fewer uninfected cells remain, each infectious virion has fewer opportunities to cause a successful cell infection and $R(t)$ decreases. Eventually the number of uninfected cells reaches a critical value, $T(t) = \Tstar$, such that
\begin{align}
&R(t) = 1 \equiv \frac{p\tau_I \cdot \gamma \beta \Tstar/\Vol}{c+\beta \Tstar/\Vol}
& \frac{\Tstar}{N_\text{cells}} = \frac{c}{(\beta N_\text{cells}/\Vol) \cdot (\gamma p\tau_I-1)} \ . \label{tstar}
\end{align}
When the number of uninfected cells remaining $T(t)$ equals $\Tstar$, then by definition $R(t)=1$ marking the infectious cell population peak, as shown in Fig~\ref{phase}(A). Thereafter, the infectious cell population declines and the number of cells that remain uninfected approaches its final value, $T(\infty)$.

For our case of interest where the initial inoculum consists of one or a few infectious virions, the MFM-predicted fraction of cells consumed, $1-T(\infty)/N_\text{cells}$, can be represented by a strictly decreasing function of $\Tstar/N_\text{cells}$ (see Eqn.\ \eqref{fsize4}). This makes sense because if the infection parameters are such that fewer cells need to be consumed by the infection to reach $R(t)=1$ (larger $\Tstar$), then fewer cells will have been consumed by the time the infection ends (larger $T(\infty)$), as shown in Fig~\ref{phase}(B). The MFM-predicted fraction of cells consumed is not, however, simply a function of $R_0$. Fig~\ref{phase}(C) shows how an infection can start with a higher $R_0$ but $R(t)$ can decrease more rapidly as the infection progresses such that $R(t)=1$ with the same fraction of uninfected cells ($\Tstar/N_\text{cells}$), ultimately resulting in the same fraction of cells consumed, $1-T(\infty)/N_\text{cells}$.

\begin{figure} 
\begin{center}
\includegraphics[width=1.0\linewidth]{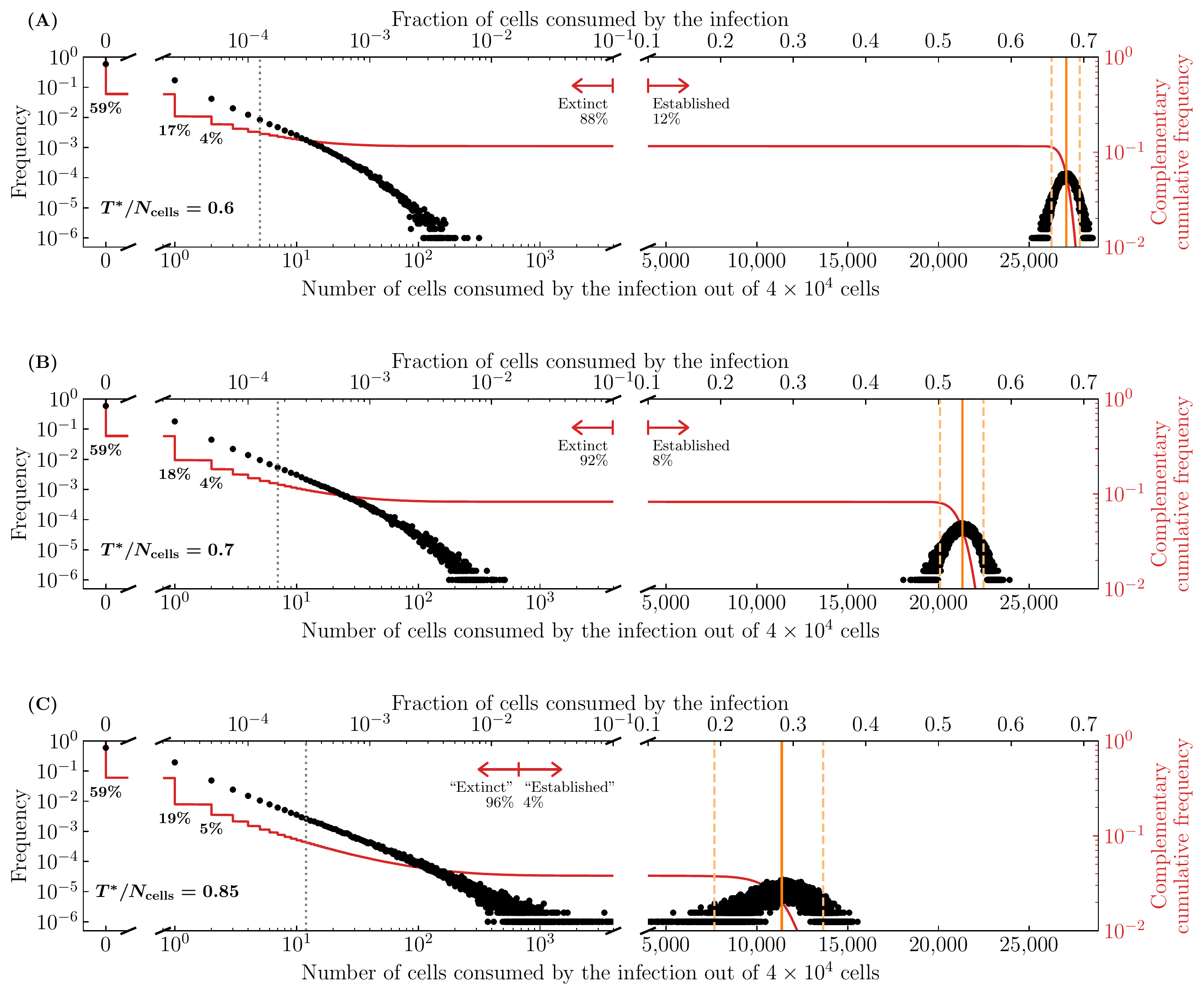}
\caption{\textbf{The effect of infection parameters on extinction and establishment.} The frequency or complementary cumulative frequency of the fraction or number of cells consumed by the infection out of $N_\text{cells} = \unit{4\times10^4}{cells}$ where the efficacy ($\varepsilon$) of an antiviral acting on the virus production rate, i.e.\ $p\to(1-\varepsilon)p$, was varied such that (A) $\Tstar/N_\text{cells} = 0.6$; (B) $0.7$ or (C) $0.85$. Everything else is generated, computed, and represented visually as described in the caption of Fig~\ref{example}.}
\label{vareps}
\end{center}
\end{figure}

Fig~\ref{vareps} shows the frequency and complementary cumulative frequency distributions of the number of cells consumed by an infection, using parameter sets where $\Tstar/N_\text{cells}$ is increasingly closer to $1$. It is achieved here by increasing the efficacy of an antiviral acting on the virus production rate, $p\to(1-\varepsilon)p$, thus increasing $\Tstar/N_\text{cells}$ (Eqn.\ \eqref{tstar}). As $\Tstar/N_\text{cells}$ approaches 1, the mean number of cells consumed by infections considered extinct increases while that by established infections decreases, and the distinction blurs as the two distributions begin to merge. In Fig~\ref{vareps}(C), the location of the red arrows demarcating so-called extinct and established infections was chosen so that $\sim$96\% of the $10^6$ DSM simulated infections, the probability of extinction given by Eqn.\ \eqref{rootm}, are to the left of the arrows. Therefore, when an antiviral is applied and the establishment probability is non-zero ($R_0>1$), varying the antiviral efficacy (hence varying $\Tstar/N_\text{cells}$) affects not only the establishment probability but also the fraction of cells consumed by both established and extinct infections. For higher antiviral efficacy (higher $\Tstar/N_\text{cells}$), the distinction between extinction and establishment becomes increasingly irrelevant.  

\begin{figure} 
\begin{center}
\includegraphics[width=1.0\linewidth]{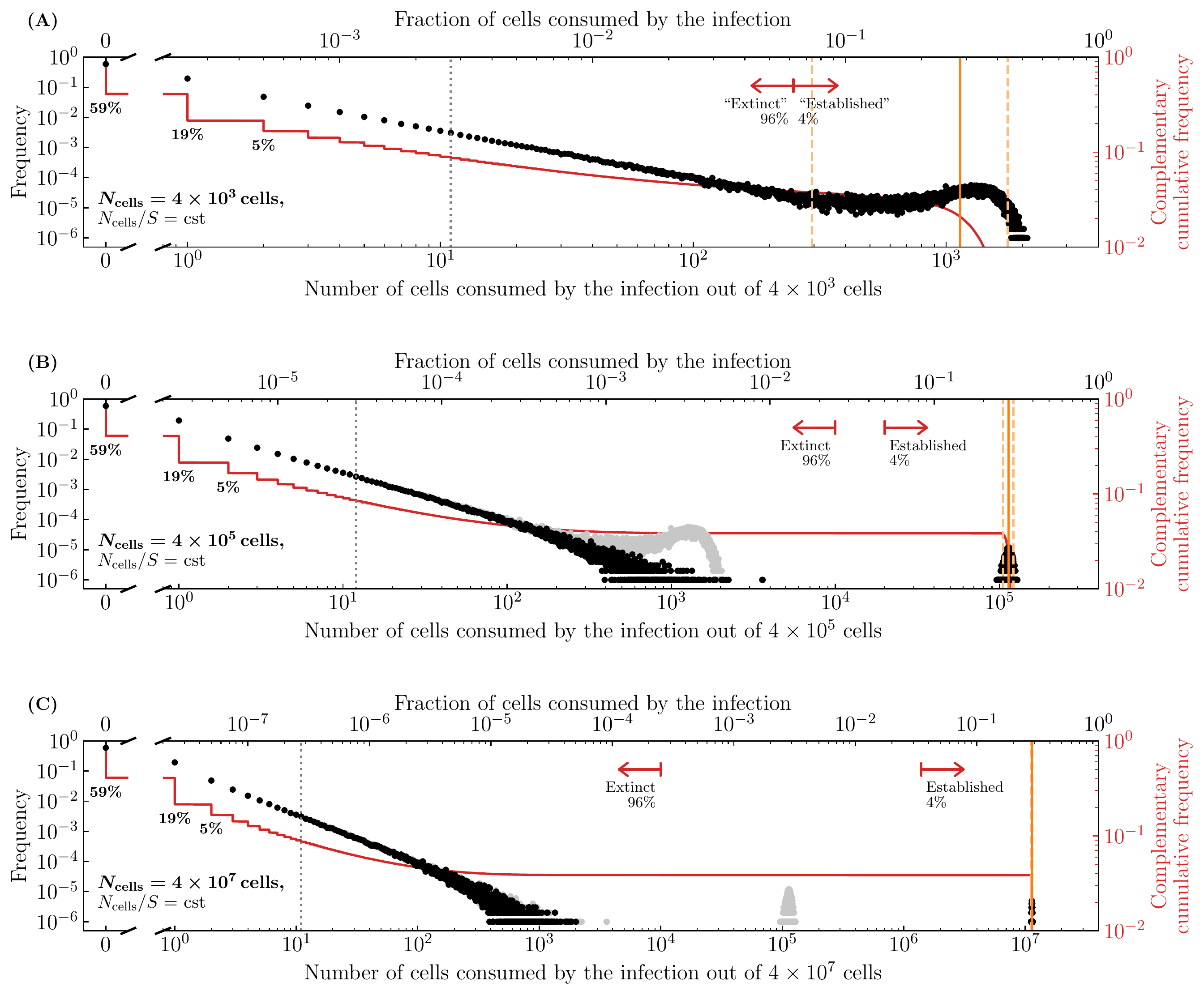}
\caption{\textbf{The effect of the size of the cell population on extinction and establishment.}
The frequency and complementary cumulative frequency of the fraction or number of cells consumed by the infection out of $N_\text{cells}$, as the latter is varied, while the cell concentration ($N_\text{cells}/\Vol$), and therefore also the establishment probability and $\Tstar/N_\text{cells}$, are fixed; (A) $N_\text{cells} = \unit{4 \times 10^3}{cells}$; (B) $\unit{4 \times 10^5}{cells}$ or (C) $\unit{4 \times 10^7}{cells}$. The frequency distribution in (A) is shown in (B) in grey, and that in (B) is shown in (C) in grey. As in Fig~\ref{vareps}(C), the efficacy ($\varepsilon$) of an antiviral acting to reduce the virus production rate, $p\to(1-\varepsilon)p$, was such that $\Tstar/N_\text{cells} = 0.85$. Everything else is generated, computed, and represented visually as described in the caption of Fig~\ref{example}.}
\label{varNx}
\end{center}
\end{figure}

Fig~\ref{varNx} explores the number of cells consumed by established and extinct infections as the size of the cell population, $N_\text{cells}$, is increased while keeping the concentration of cells ($N_\text{cells}/\Vol$) and thus the establishment probability and $\Tstar/N_\text{cells}$ fixed. In Fig~\ref{varNx}(A) with the smallest cell population considered, the distributions for the number of cells consumed by established versus extinct infections overlap.

In Fig~\ref{varNx}(B,C), as the size of the cell population ($N_\text{cells}$) is increased, the distribution for the number of cells consumed by extinct infections (left side of the graph) is unaffected, and therefore the fraction of cells consumed becomes increasingly small. The 95\% CI upper bound remains fixed at $\sim\unit{11}{cells}$ which corresponds to 0.3\%, 0.003\% and 0.00003\% out of $N_\text{cells}$ as the latter is increased from \unit{4\times10^3}{cells} to \unit{4\times10^7}{cells}. In contrast, in the established infections, the number of cells consumed increases as $N_\text{cells}$ increases, while the median fraction of cells consumed remains unchanged, $\sim28\%$. Therefore, as the size of the cell population is increased, there is a greater distinction between extinction and establishment. This means that, with a larger cell population, a higher antiviral efficacy (higher $\Tstar/N_\text{cells}$) is required for the distinction between extinction and establishment to vanish.

\subsection{Reduction in the number of cells consumed by infections under antiviral therapy}

\begin{figure} 
\begin{center}
\includegraphics[width=0.7\linewidth]{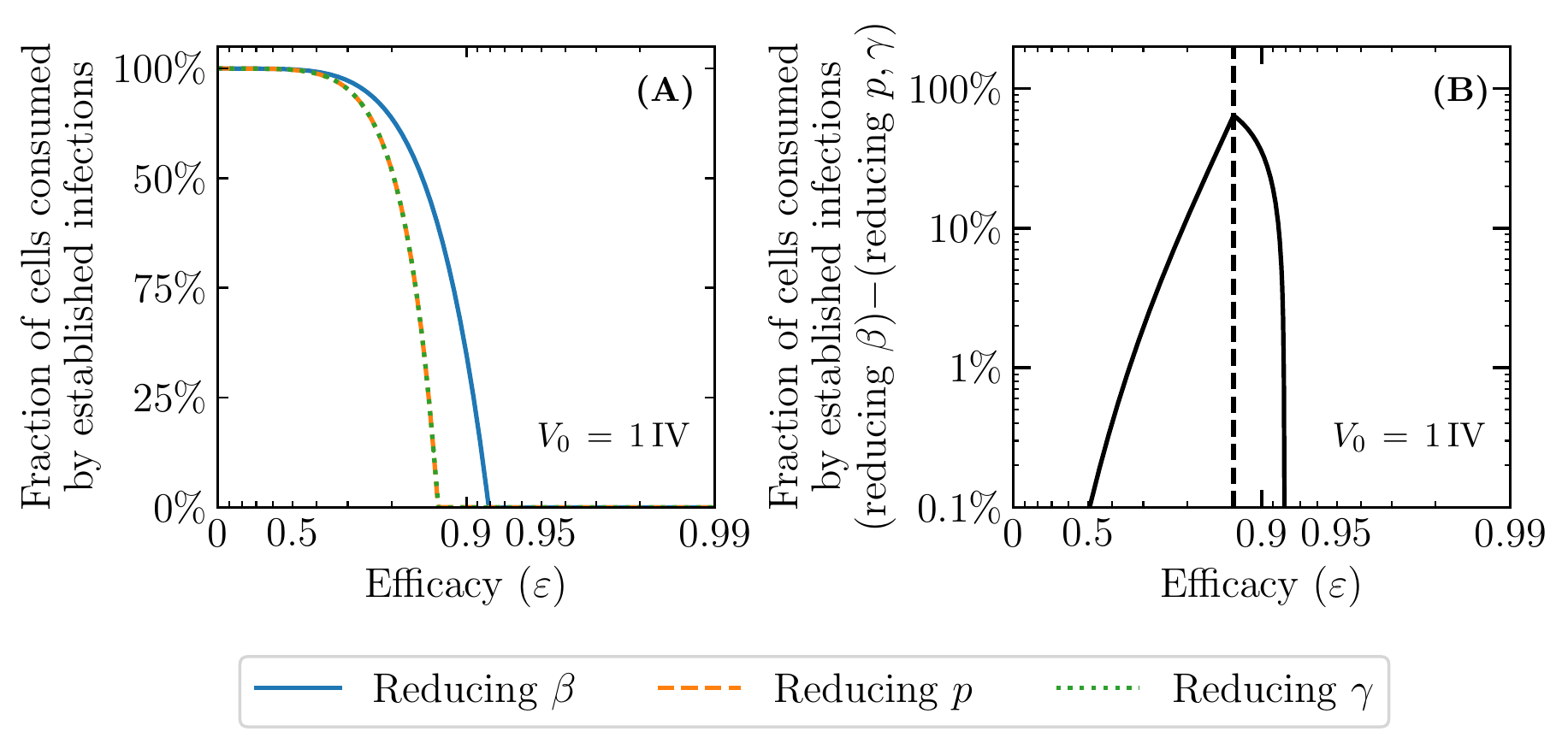}
\caption{\textbf{Effect of different antivirals on the MFM-predicted fraction of cells consumed by established infections.}
(A) The MFM-predicted fraction of cells consumed by established infections initiated with only one infectious virion as a function of efficacy ($\varepsilon$) for antivirals acting prophylactically either to reduce the virus entry rate, $\beta\to(1-\varepsilon)\beta$ (blue solid line), the virus production rate, $p\to(1-\varepsilon)p$ (orange dashed line), or the probability of a successful cell infection post viral entry, $\gamma\to(1-\varepsilon)\gamma$ (green dotted line).
(B) The MFM-predicted fraction of cells consumed by established infections shown in (A) for an antiviral acting to reduce $\beta$ minus that for one acting to reduce either $p$ or $\gamma$. The dash line in (B) indicates the value of efficacy where the fraction of cells consumed for an antiviral acting on $p$ or $\gamma$ equals zero. Infection parameters are the same as in Fig~\ref{mrhoV}.} 
\label{figfsize}
\end{center}
\end{figure}

Prophylactic antiviral therapy not only reduces an infection's establishment probability, $\rhoV$ in Eqn.\ \eqref{rootm}, but as seen above it also decreases the number of cells consumed by established infections while increasing that consumed by extinct infections (see Fig~\ref{vareps}). The median of the distribution for the number of cells consumed by established infections in the DSM simulations approximately corresponds to the MFM-predicted value, $N_\text{cells}-T(\infty)$. Let us then explore the effect of antivirals on the MFM-predicted fraction of cells consumed by an infection.

Fig~\ref{figfsize}(A) shows the MFM-predicted fraction of cells consumed by an established infection, given an initial inoculum of only one infectious virion, as a function of antiviral efficacy ($\varepsilon$) for antivirals acting on $\beta$, $p$, or $\gamma$. Since an antiviral acting on $p$ or $\gamma$ have the same effect on $\Tstar = (c\cdot\Vol)/[\beta \cdot (\gamma p \tau_I -1)]$ (see Eqn.\ \eqref{tstar}), they also cause the same MFM-predicted fraction of cells to be consumed by established infections, $1-T(\infty)/N_\text{cells}$ (see Methods, Section \ref{final-size-section}). On the other hand, an antiviral acting on $\beta$, for the same $\varepsilon$, results in a smaller $\Tstar$ and thus larger MFM-predicted fraction of cells consumed. Fig~\ref{figfsize}(B) illustrates how these differences, and therefore their biological relevance, is highly dependent on the antiviral efficacy under consideration. For example, the difference in the MFM-predicted fraction of cells consumed by established infections for antivirals acting on $\beta$ rather than $p$ or $\gamma$ is $<$0.1\% at an antiviral efficacy $\varepsilon<0.5$ or $>0.92$.

\begin{figure} 
\begin{center}
\includegraphics[width=1.0\linewidth]{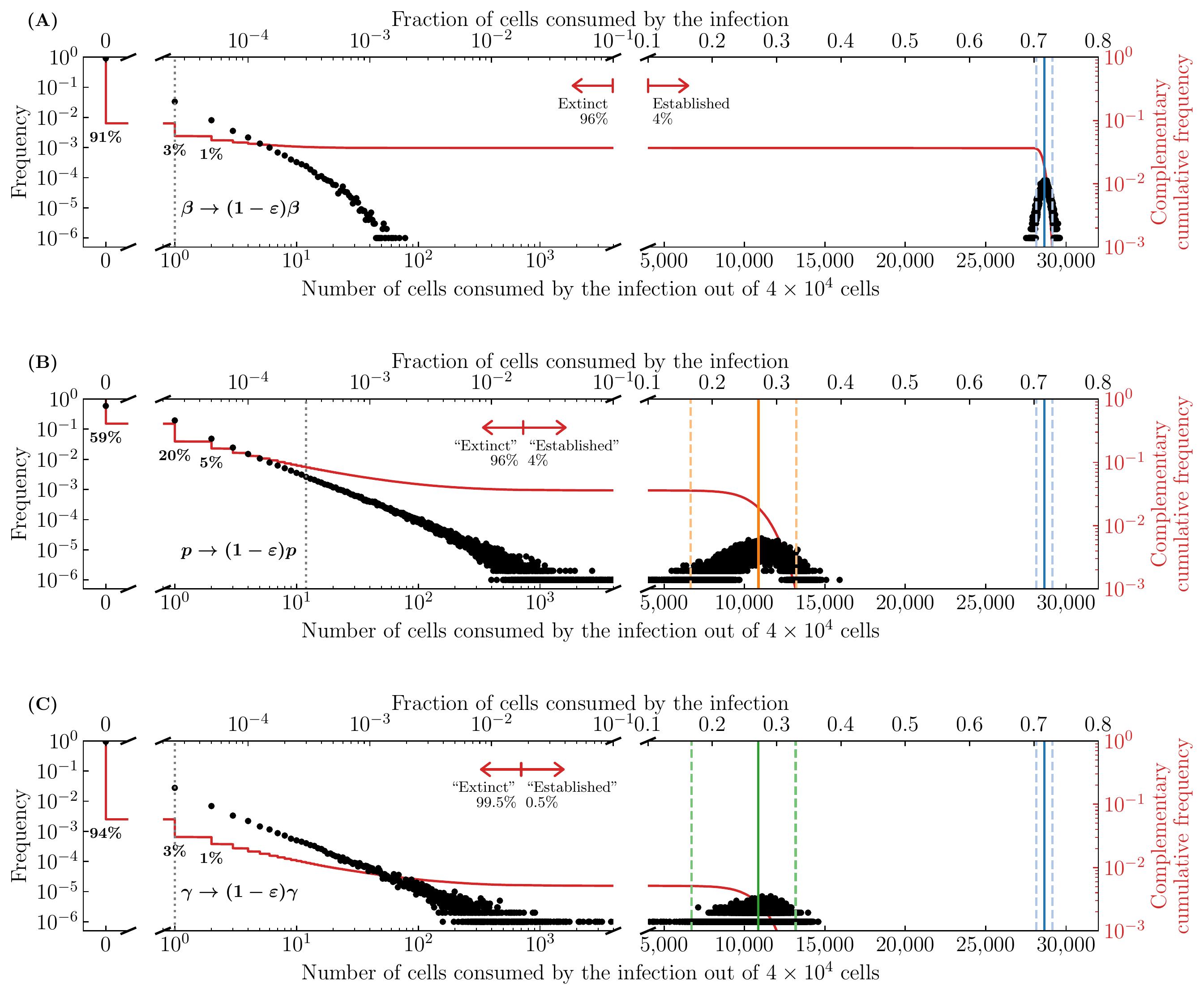}
\caption{\textbf{Antiviral with different modes of action cause different fraction of cells to be consumed by infections.}
The number of cells consumed by the infection out of $N_\text{cells} = \unit{4 \times 10^4}{cells}$ for antivirals with efficacy $\varepsilon=0.86$ acting either to reduce (A) the virus entry rate, $\beta\to(1-\varepsilon)\beta$, (B) the virus production rate, $p\to(1-\varepsilon)p$, or (C) the probability of a successful cell infection post viral entry, $\gamma\to(1-\varepsilon)\gamma$. Antiviral modes of action are represented by different colours: reducing $\beta$ (blue), $p$ (orange) or $\gamma$ (green). Everything else is generated, computed, and represented visually as described in the caption of Fig~\ref{example}.}
\label{exfsize}
\end{center}
\end{figure}

Fig~\ref{exfsize} shows the number of cells consumed by extinct and established infections, given an initial inoculum of a single infectious virion, for antivirals acting on $\beta$, $p$ or $\gamma$ at an antiviral efficacy of $\varepsilon = 0.86$. This efficacy was chosen so that, given the infection parameters, the establishment probability is the same for an antiviral acting on $\beta$ or on $p$ ($\rhoV = 4\%$), yet it corresponds to a lower establishment probability for an antiviral acting on $\gamma$ ($\rhoV=0.5\%$). This can be seen from Fig~\ref{mrhoV}: it corresponds to the efficacy where the orange line (antiviral acting on $p$) crosses the blue line (antiviral acting on $\beta$).

At this efficacy, and for the chosen infection parameters, biologically and statistically significantly more cells are consumed by established infections for an antiviral acting on $\beta$ than on $p$ or $\gamma$. For extinct infections, which are a more likely outcome, the least number of cells are consumed for an antiviral acting on $\beta$, followed by $\gamma$, and then $p$. When averaging over all infection outcomes, i.e.\ extinct and established infections, an antiviral acting on $\gamma$ will lead to an average of 56 cells consumed, followed by one acting on $p$ with 396 cells, and finally on $\beta$ with 1,055 cells. Rather than looking at extinct versus established infections, one might consider that triggering an immune response, and therefore possibly also a wide range of symptoms, is the most important aspect of infection control. This could conceivably be triggered once a certain number of cells, e.g.\ $\sim20$ or so, have been consumed by infection. In this case, an antiviral acting on $\gamma$, more so than one acting on $\beta$, followed by one acting on $p$, would help avoid triggering an immune response, i.e.\ have a smaller proportion of infection outcomes where less than $\sim20$ cells are consumed.

We investigated a second parameter set also explored in \czuppon, which is characterized by a 10-fold decrease in the number of cells $N_\text{cells}$, a 10-fold increase in the virus production rate $p$ (hence, of the average burst size $\B = p \tau_I$), and a corresponding $\sim$10-fold decrease in the virus entry rate $\beta N_\text{cells}/\Vol$ (see Supplementary Material \ref{highN-section}). For this second parameter set, we found that an antiviral acting on $\gamma$ was comparable to one acting on $\beta$, but better than one acting on $p$, for reducing the establishment probability. All 3 modes of action had a similar effect on the MFM-predicted fraction of cells consumed by established infections. As such, the degree to which an antiviral acting on $\gamma$ is better to reduce the establishment probability than other antivirals, or an antiviral acting on $\beta$ will leave more cells consumed by established infections, will depend on infection parameters. It is therefore critical that such parameters be well-determined if such investigations are to yield meaningful predictions.

Lastly, as in \czuppon, we also considered antiviral therapy for an infection initiated with only one infectious cell which could be representative of post-exposure antiviral therapy (see Supplementary Material \ref{PET-section}). In this case, we found that an antiviral acting on $p$ or $\gamma$ have the same effectiveness, greater than one acting on $\beta$, in reducing the establishment probability. There was no meaningful difference in the MFM-predicted fraction of cells consumed by established infections for infections initiated with one infectious cell rather than one infectious virions.

Overall, these results over different infection parameters and antiviral efficacy indicates that, under certain conditions, the fraction of cells consumed by both or either extinct and established infections can be an important consideration when evaluating and comparing antivirals with different modes of action. It also highlights the importance of properly identifying biologically relevant base parameter values in order to provide meaningful comparisons.

\subsection{Continuous versus burst release of virus and the impact of the virus burst size distribution}

\begin{figure}
\begin{center}
\includegraphics[width=1.0\linewidth]{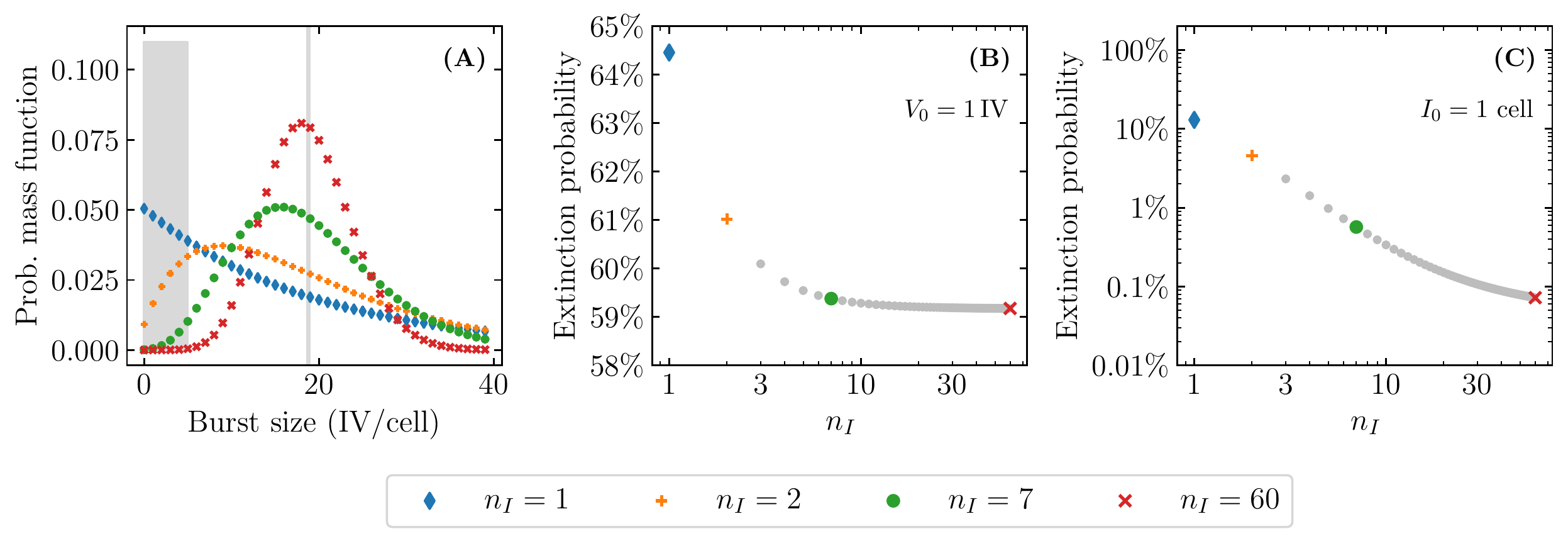}
\end{center}
\caption{\textbf{Effect of the infectious phase duration's shape parameter, $n_I$, on the extinction probability.}
(A) Negative binomial-distributed virus burst size as a function of the shape parameter ($n_I$) of the Erlang-distributed infectious cell lifespan. The shaded region highlight a region where the burst size is small ($\le5$ infectious virions) and therefore more likely to result in infection extinction. The vertical grey line corresponds to the average burst size value ($\B = p \tau_I = \unit{18.8}{\IU/cell}$), which is constant as $n_I$ is varied.
(B,C) The probability of infection extinction as a function of $n_I$, given an infection initiated with only one infectious virion (B); or one infectious cell (C). Note that the extinction probability is shown on a linear scale in (B) but a logarithmic scale in (C) to better capture the relationships.
Unless otherwise specified, the parameters are the same as in Fig~\ref{mrhoV}.} 
\label{nI-effect}
\end{figure}

In the MFM and DSM used herein, virus is released continuously at a fixed rate of $p$ infectious virion per hour by infectious cells. In the DSM, this fixed rate maps to a Poisson-distributed, discrete, stochastic number of infectious virions produced per time step by each infected cell over the duration of its infectious lifespan. The duration of this infectious phase is represented by an Erlang distribution characterized by shape parameter $n_I$ and average duration $\tau_I$. As shown in Eqn.\ \eqref{burstr}, this means that the DSM's stochastic virus burst size follows a negative binomial or Pascal distribution with a mean corresponding to the MFM's virus burst size, $\B = p\tau_I$, where $n_I$ now corresponds to the distribution's integer-valued, stopping-time parameter. Fig~\ref{nI-effect}(A) illustrates how the DSM's burst size distribution varies as a function of $n_I$, for a fixed average burst size.

Up to this point, as with most DSMs, the duration of the infectious phase has been assumed to be exponentially-distributed ($n_I = 1$ in Eqn.\ \eqref{mfeq}). This choice does not affect the MFM-predicted fraction of cells consumed by established infections, since $n_I$ does not appear in Eqn.\ \eqref{fsize3}, but it does affect the extinction probability, as per Eqn.\ \eqref{rootm}. Previously, \yan have shown that increasing $n_I$ leads to a decrease in the extinction probability when infection is initiated with a small initial number of infectious virions. Fig~\ref{nI-effect}(B) shows this for the infection parameters used so far for an infection initiated with a single infectious virion.

Recall that the extinction probability is a recursive expression with 2 main terms,
\begin{align*}
\rhoV = \qfail + (1-\qfail) \cdot \sum_{m=0}^\infty \underbrace{\text{NB}\left(m\left|r_\text{fail}=n_I,p_\text{success}=\frac{\B}{n_I+\B}\right.\right)}_{\text{burst size distribution}} \cdot \left(\rhoV\right)^m \ .
\end{align*}
where the first term, $\qfail$, is the probability that the initial virion inoculum fails to productively infect a single cell, and the second is the likelihood that it does actually cause a cell infection which then itself fails to establish an infection. Fig~\ref{nI-effect}(C) shows how the contribution from the second term, i.e.\ the extinction probability for an infection initiated with a single infected cell, decreases as $n_I$ increases. What is observed is Fig~\ref{nI-effect}(B), therefore, is the shrinking contribution to $\rhoV$ by the second term as $n_I$ increases.

This second term depends critically on the burst size distribution which, in our DSM, is a result of the Erlang-distributed infectious phase duration. As the latter distribution goes from an exponential distribution ($n_I=1$) to a fat-tailed ($n_I\in[2,7]$), to a normal-like ($n_I=60$), tending ultimately to a Dirac delta distribution ($n_I\to\infty$), it becomes less probable that an infectious cell will have a very small burst size, as shown in Fig~\ref{nI-effect}(A). For example, the probability that an infectious cell will have a burst size less than or equal to 5 infectious virions (shaded region in Fig~\ref{nI-effect}(A)) is 27\% when $n_I=1$ compared to 0.07\% when $n_I=60$ ($>380\times$ less likely). This is why, as $n_I$ increases, the probability of infection extinction once one cell has been infected decreases, and this effect is more pronounced for larger viral burst sizes.

\begin{figure}
\begin{center}
\includegraphics[width=1.0\linewidth]{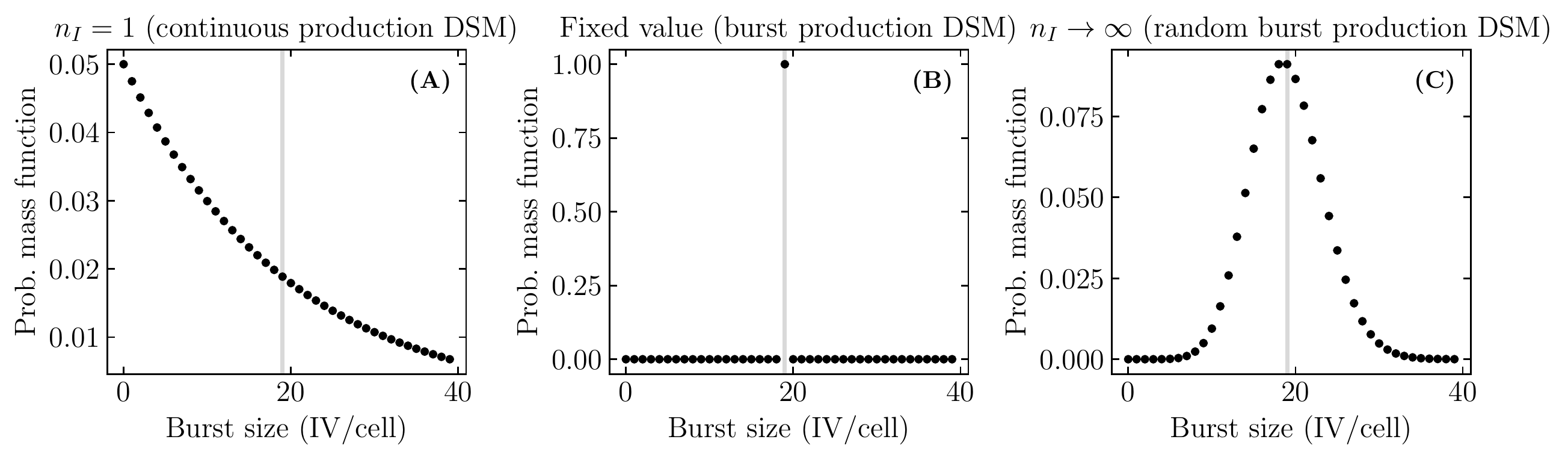}
\end{center}
\caption{\textbf{Visual representation of the burst size distributions explored in \pearson.} Burst size distribution for (A) $n_I=1$ (so-called continuous production in \pearson); (B) a fixed value (so-called burst production in \pearson); (C) $n_I \to \infty$ (so-called random burst production in \pearson). The grey line represents the average burst size value $\B = \unit{19}{\IU/cell}$.}
\label{pearson}
\end{figure}

This brings up a misconception, introduced by \pearson, that has been repeatedly stated by others \cite{yuan11,shaw12,xiao13,sanchez15}. \pearson derived the extinction probability for different DSMs: DSMs where infectious virions are continuously released at a fixed rate over an infectious cell's lifespan (so-called continuous production); and DSMs where infectious virions are released all at once, as a single burst, upon an infected cell's death (so-called burst production). In their work, \pearson conclude that continuous production is more likely to lead to infection extinction than burst production. But since each infection event, i.e.\ whether or not each infectious virion produced results in an infection, can be treated independently based on the models under consideration, the timing of the release of each infectious virions cannot and indeed does not affect the extinction probability.

The difference observed by \pearson in the extinction probabilities for the DSMs explored is entirely a consequence of the different burst size distributions used by the authors in the different DSMs, and not their mode of release (continuous versus burst), as stated therein. The burst size for their continuous production DSM is geometrically-distributed \cite{czuppon21}, as depicted in Fig~\ref{pearson}(A), where the geometric distribution is the discrete analogue of the exponential distribution ($n_I=1$ in our DSM). The burst size for their burst production DSM is a fixed value \cite{pearson11}, shown in Fig~\ref{pearson}(B), and that for their random burst production DSM is Poisson-distributed \cite{pearson11}, depicted in Fig~\ref{pearson}(C). If their burst production DSM is modified so that the burst size distribution follows the same geometric distribution as that for the continuous production DSM, both DSMs have the same extinction probability. With our DSM, the burst size is geometrically-distributed when $n_I = 1$ and Poisson-distributed when $n_I \to \infty$ (see Methods, Section \ref{qprod-section}). As such, the comparison of the extinction probability for the so-called continuous versus random burst production DSMs in \pearson is in fact a comparison of our continuous production DSM with two extreme values of $n_I$, i.e.\ $n_I = 1$ vs.\ $n_I \to \infty$. In other words, what \pearson actually investigated was the decrease in the extinction probability as $n_I$ increases.

Having explored how the shape parameter of the infectious phase duration distribution ($n_I$) affects the extinction probability, we can now consider its impact in evaluating and comparing antivirals. For a higher value of $n_I$, since the extinction probability is lower, the establishment probability, or $1-$(extinction probability), is higher. This means that although the effectiveness of antivirals in reducing the establishment probability does not change qualitatively (the better ones remain better), it does change quantitatively (see Supplementary Material Fig~\ref{mrhoVnI}). For example, the choice of $n_I$ affects $\varepsilon_{50}$, the efficacy at which the establishment probability is 50\% of its value without antivirals, i.e.\ at $\varepsilon = 0$ (see Table \ref{eps50}). \czuppon state that for an infection initiated with 10 infectious virions, $\varepsilon_{50}$ is 81\% for an antiviral acting on $\beta$, and 85\% for an antiviral acting on $p$, where the authors have assumed $n_I=1$. When $n_I = 60$, however, we find that $\varepsilon_{50}$ is comparable for an antiviral acting on $\beta$ vs.\ $p$ (85\% vs.\ 86\%), and far less (77\%), and therefore potentially easier to achieve, for one acting on $\gamma$. This again highlights the importance of properly estimating parameters before making quantitative comparisons of antiviral regimens.

\begin{table*}
\begin{center}
\caption{Efficacy at which the establishment probability is 50\% of its value without antivirals ($\varepsilon_{50}$).}
\label{eps50}
\begin{tabular}{|ccc|} 
\hline
Mode of action & \multicolumn{2}{c|}{Actual $\varepsilon_{50}$ (relative to $\gamma$)} \\
& for $n_I =1$ & for $n_I = 60$ \\
\hline
\multicolumn{3}{|c|}{for an initial inoculum of $V_0=1\,$\IU} \\
\hline
Reducing $\gamma$ & 43\% (0\%) & 55\% (0\%) \\
Reducing $\beta$ & 57\% (14\%) & 67\% (12\%) \\
Reducing $p$ & 77\% (34\%) & 83\% (28\%) \\
\hline 
\multicolumn{3}{|c|}{for an initial inoculum of $V_0=10\,$\IU} \\
\hline
Reducing $\gamma$ & 71\% (0\%) & 77\% (0\%) \\
Reducing $\beta$ & 81\% (10\%) & 85\% (8\%) \\
Reducing $p$ & 85\% (14\%) & 86\% (9\%) \\
\hline
\end{tabular}
\end{center}
\end{table*}

%%%%%%%%%%%%%%%%%%%%%%%%%%%%%%%%%%%%%%%%%%%%%%%%%%%%%%%%%%%%%%%%%%%%%%%%%%%%%%%%
%%%%%%%%%%%%%%%%%%%%%%%%%%%%%%%%%%%%%%%%%%%%%%%%%%%%%%%%%%%%%%%%%%%%%%%%%%%%%%%%
%%%%%%%%%%%%%%%%%%%%%%%%%%%%%%%%%%%%%%%%%%%%%%%%%%%%%%%%%%%%%%%%%%%%%%%%%%%%%%%%
%%%%%%%%%%%%%%%%%%%%%%%%%%%%%%%%%%%%%%%%%%%%%%%%%%%%%%%%%%%%%%%%%%%%%%%%%%%%%%%%

\clearpage
\section{Discussion}

Discrete, stochastic mathematical models (DSMs) of viral infection kinetics usually do not represent failure of a virion to cause an infection post cell entry. Yet biologically, there are many way in which a virion, post cell entry, will fail to cause a cell infection that will yield infectious progeny. Herein, we constructed a DSM of viral infection kinetics with an explicit parameter ($\gamma$) to represent the probability that a virion will cause a productive infection post cell entry. The DSM was first used to estimate the extinction probability of an infection, i.e.\ the probability that an infection will fail to take hold or spread significantly. 

Previously, \czuppon evaluated prophylactic antivirals acting on viral entry ($\beta$) or production ($p$) to reduce the establishment probability, or $1-$(extinction probability), for a SARS-CoV-2 infection. Extending this work, we investigated how an antiviral acting on $\gamma$ would compare against antivirals acting on $\beta$ or $p$. We found that a prophylactic antiviral acting on $\gamma$ was best at reducing the establishment probability when infection is initiated with a small number of infectious virions. When instead an infection is initiated with an initial number of infectious cells, possibly representative of post-exposure antiviral therapy, we found that an antiviral acting on $\gamma$ or $p$ caused the same reduction in the establishment probability, better than that for an antiviral acting on $\beta$. More generally, we found that the degree to which an antiviral with a particular mode of action is better than another critically depends on the chosen infection parameters.

In HIV antiviral therapy, reverse transcriptase inhibitors (RTIs) prevent the transcription of viral DNA from viral RNA, a step that occurs after viral entry and is necessary for viral replication. Conway et al.\ \cite{conway13} have reported that, under pre-exposure antiviral therapy, RTIs which they represented as acting on their DSM's cell infection rate ($\beta$) are more effective than protease inhibitors (acting on the virus production rate, $p$), at reducing the risk of HIV infection. Using our DSM, we can show that an antiviral acting on $\gamma$, which better captures the mode of action of a RTI, is even more effective than one acting on $\beta$ at reducing the risk of infection for pre-exposure antiviral therapy, at least for some of the infection parameter sets explored in Conway et al.\ \cite{conway13} (Supplementary Material Fig~\ref{conway13test}(A)). These findings echo the findings reported herein for SARS-CoV-2 parameters, as to the equal or better performance of antivirals acting to reduce $\gamma$, compared to $p$ or $\beta$. The introduction of parameter $\gamma$ to capture productive cell infection failure \emph{after} an infectious virion has successfully entered a cell is an important consideration when comparing antivirals based on their mode of action.

We found that the distribution for the number of cells consumed by DSM-simulated infections tended to fall into two patterns: those that consumed a low vs a high number of cells, which one could identify as extinct and established infections, respectively. We found that the fraction of infections resulting in a low number of cells consumed matched the infection extinction probability derived for our DSM, which depends on the infection parameters. We also found that the median number of cells consumed by established infections closely matched that predicted by the mean-field mathematical model (MFM), also expressed in terms of the infection parameters. Therefore, as the efficacy of an antiviral acting on a particular infection parameter is increased, it will decrease both the infection's establishment probability and the number of cells consumed by infections identified as established. This finding had 2 important implications. Firstly, since the infection's establishment probability and the number of cells consumed by established infections depend differently on the DSM's parameters, antivirals acting on different parameters can decrease one quantity more effectively than the other. Secondly, and perhaps more importantly, as antiviral efficacy increases, the distinction in the number of cells consumed by established and extinct infections vanishes. For example, at a drug efficacy that yields an equal probability of infection extinction, we found that an antiviral acting on $\beta$, compared to one acting on $p$, resulted in $\sim3\times$ more cells consumed by so-called established infections. At that same efficacy, an antiviral acting on $\gamma$ yielded both a lower probability of infection establishment, and resulted in fewer cells consumed overall. Under such conditions, the average number of cells consumed by all infections, established and extinct, becomes a more biologically relevant quantity to track than the probability of infection extinction. Looking at the probability that infections will consume more than some biologically critical number of cells, e.g.\ that identified as sufficient to trigger an immune response and its associated symptoms or minimally sufficient for transmission, could be more appropriate.

To facilitate direct comparison of our work herein to past work by others, most of our results were based on an infectious period, during which an infected cell is releasing virus, whose duration is exponentially distributed (Erlang-distributed with shape parameter $n_I=1$). But the infectious period for many different viruses has been shown in vitro to be inconsistent with an exponential distribution, and instead follow a more normal-like distribution ($n_I>\sim10$). \yan have shown previously that increasing $n_I$ leads to a decrease in the extinction probability. Herein we established that this happens because the probability that an infectious cell will stochastically produce a very small burst size becomes vanishingly small as $n_I$ increases. We showed that while $n_I$ does not affect the fraction of cells consumed by established infections, increasing $n_I$ decreases the probability of infection extinction, requiring higher antiviral efficacy to achieve the same risk reduction. Importantly, we demonstrated that continuous release of virus over time results in the same probability of infection extinction as virus released at once as a burst, addressing a misconception introduced in \pearson, and since repeated \cite{yuan11,shaw12,xiao13,sanchez15}. In the end, the differences observed by \pearson were a consequence of the different burst size distributions used in comparing continuous ($n_I=1$) and random burst ($n_I=\infty$) production, rather than a consequence of the timing of viral release.

The DSM introduced herein is general and, like the ODE model on which it is based, it should be applicable to a wide range of different viruses. A logical improvement to this DSM would be the inclusion of semi-infectious and defective interfering particles. Our DSM explicitly represents virion that successfully enter a cell but fail to cause an infection. Our DSM assumes this process leaves the cell in the same state as if entry had not occurred. Biologically, repeated viral failure post cell entry would likely trigger antiviral pathways within the cell, leaving it in a different state than its naive, susceptible state. Additionally, the accumulation of failed virions, if it occurs over a sufficiently short time, could eventually add up to one functional productive infection or lead to defective interfering particle production. The resulting, more complicated DSM would have more parameters to identify, and would require a very rich, plentiful data set to validate and parametrize. This poses a challenge that could be hard to overcome.

%%%%%%%%%%%%%%%%%%%%%%%%%%%%%%%%%%%%%%%%%%%%%%%%%%%%%%%%%%%%%%%%%%%%%%%%%%%%%%%%
%%%%%%%%%%%%%%%%%%%%%%%%%%%%%%%%%%%%%%%%%%%%%%%%%%%%%%%%%%%%%%%%%%%%%%%%%%%%%%%%
%%%%%%%%%%%%%%%%%%%%%%%%%%%%%%%%%%%%%%%%%%%%%%%%%%%%%%%%%%%%%%%%%%%%%%%%%%%%%%%%

\clearpage
\section{Methods}

\subsection{Parameter values used to generate all figures}
\label{summary-params}

The following are the parameters used to generate each figure found within the manuscript.

\begin{description}

\item[Figs~\ref{mrhoV}, \ref{figfsize}, \ref{1fail-fig}, \ref{burst-size-fig}, \ref{rhoV-fig}]
	$p = \unit{(11.2/24)}{\IU/(cell \cdot h)}$, $N_\text{cells} = \unit{4\times 10^4}{cells}$, $R_0 = 7.69$,  $\tau_E = \unit{24/5}{h}$, $\tau_I = \unit{(24/0.595)}{h}$, $c = \unit{(10/24)}{h^{-1}}$, $n_E = 1$, $n_I = 1$ and $\beta = [cR_0/\tau_I]/[N_\text{cells}(p-R_0/\tau_I)]$ \cite{czuppon21}. Also, $\gamma = \unit{1}{cell/\IU}$ and $S = \unit{1}{mL}$. 

\item[Fig~\ref{example}]
	As in Fig~\ref{mrhoV} but with $p \to (1-\varepsilon)p$, where $\varepsilon=0.79$.

\item[Fig~\ref{phase}(A)]
	As in Fig~\ref{mrhoV} but with $\beta \to (1-\varepsilon)\beta$, where $\varepsilon = 0.81$.

\item[Fig~\ref{vareps}]
	As in Fig~\ref{example} but with $\varepsilon = 0.82,\,0.84,\,0.86$.

\item[Fig~\ref{varNx}]
	As in Fig~\ref{example} but with $\varepsilon = 0.86$, $N_\text{cells} = \unit{4 \times [10^3,10^5,10^7]}{cells}$ and $N_\text{cells}/\Vol$ fixed.

\item[Fig~\ref{exfsize}]
	As in Fig~\ref{mrhoV} but with an antiviral acting either on $\beta$, $p$ or $\gamma$, where $\varepsilon = 0.86$.

\item[Fig~\ref{nI-effect}]
	As in Fig~\ref{mrhoV} but with $n_I = 1,\,2,\,7,\,60$.

\item[Fig~\ref{pearson}]
	$\B = \unit{19}{\IU/cell}$.

\end{description}
Fig~\ref{phase}(B,C) and Fig~\ref{Tstar} are generated using Eqn.\ \eqref{phaseeq} and Eqn.\ \eqref{fsize4} respectively, neither of which depend directly on infection parameters.

%\begin{table*}[h]
%\begin{center}
%\caption{Parameter set from \czuppon.}
%\label{params}
%\begin{tabular}{llllll}
%$p$ [\IU/($\text{cell} \cdot \text{h})$] & $N_\text{cells}$ [cells] & ${R_0}^*$ & $\tau_I$ [h] & $c$ [h$^{-1}$] & $n_I$ \\
%\hline
%11.2/24 & $4 \times 10^4$ & 7.69 & 24/0.595 & 10/24 & 1 \\
%\hline
%\end{tabular}\\[0.5em]
%\begin{minipage}{0.7\linewidth}
%* $\beta = [cR_0/\tau_I]/[N_\text{cells}(p-R_0/\tau_I)]$ 
%\end{minipage}
%\end{center}
%\end{table*}

\subsection{MFM using the expected value of the DSM's random variables} \label{mf-solver-expected-val}

One way of defining a MFM is by replacing the random variables of the DSM by their expected values, listed in Table \ref{mean}. This is trivial for the random variables that are drawn from well-defined probability distributions. But it is not for $N^\text{inf}$, which is determined using a \texttt{Python} expression that simulates randomly placing $V^\text{suc}$ infectious virions into $T^t$ cells chosen at random with replacement, \texttt{numpy.random.choice}, and counting the number of distinct cells that received one or more infectious virions, \texttt{len(numpy.unique(...))}.

Consider then $e_n$, the expected number of empty cells after placing $n$ infectious virions into $T^t$ cells. After placing $n-1$ infectious virions into $T^t$ cells, there is on average $e_{n-1}$ empty cells by definition. There is then a $e_{n-1}/T^t$ probability to place the last infectious virion into an empty cell. It follows that
\begin{align}
e_n &= \frac{e_{n-1}}{T^t}(e_{n-1}-1) + \bigg(1-\frac{e_{n-1}}{T^t}\bigg) e_{n-1} \nonumber \\
e_n &= \frac{e^2_{n-1}}{T^t}-\frac{e_{n-1}}{T^t}+e_{n-1}-\frac{e^2_{n-1}}{T^t}  \nonumber \\
e_n &= e_{n-1} \bigg(\frac{T^t-1}{T^t} \bigg) 
\end{align}
By recursion and since trivially $e_0\,=\,T^t$,
\begin{align}
e_n &= T^t \bigg(\frac{T^t-1}{T^t} \bigg)^n
\end{align}
The expected number of $N^\text{inf}$ cells after placing $V^\text{suc}$ infectious virions into $T^t$ cells is therefore represented as
\begin{align}
T^t \Bigg[1-\bigg(\frac{T^t-1}{T^t} \bigg)^{V^\text{suc}} \Bigg]
\end{align}

\begin{table*}[h]
\begin{center}
\caption{Expected value of the random variables used by the DSM.}
\label{mean}
\begin{tabular}{ll}
Random variable & Expected value \\
\hline
{$E_i^\text{out}$} & $E_i^t \cdot \Delta tn_E/\tau_E$ where $i=1,2,...,n_E$\\
{$I_j^\text{out}$} & $ I_j^t \cdot \Delta tn_I/\tau_I$ where $j=1,2,...,n_I$ \\
{$V^\text{prod}$} & $\Delta t\cdot p\,\sum_{j=1}^{n_I} I_j^t$ \\
{$V^\text{decay},V^\text{enter},V^\text{remain}$} & $V^t \cdot \Delta tc,V^t \cdot \Delta t\beta_\text{} T^t/\Vol, V^t \cdot (1-\Delta tc-\Delta t\beta_\text{} T^t/\Vol)$ \\
{$N^\text{inf}$} & $T^t\big[1-[(T^t-1)/T^t]^{V^\text{suc}} \big]$ for $T^t \geq 1$ and 0 for $T^t < 1$ \\
& \hspace{2em} where the expected value of $V^\text{suc}$ is $V^{\text{enter}} \cdot \gamma$ \\
\hline
\end{tabular}
\end{center}
\end{table*}

\subsection{MFM-predicted fraction of cells consumed by the infection} \label{final-size-section}

Let us derive the MFM-predicted fraction of cells consumed by the infection given that the infection starts with a number of infectious virions $V(0)$. Following \cite{luo12}, we have the following relations from Eqn.\ \eqref{mfeq},
\begin{align}
\dot{T}+\sum_{i=1}^{n_E} \dot{E}_i+\sum_{j=1}^{k}\dot{I}_j = -n_II_k/\tau_I &\Longrightarrow I_k = -(\tau_I/n_I)\left[\dot{T}+\sum_{i=1}^{n_E} \dot{E}_i+\sum_{j=1}^{k}\dot{I}_j\right] \\
\frac{\text{d}}{\text{d}t} \ln(T) = \frac{\dot{T}}{T} = -\gamma \beta V/\Vol &\Longrightarrow \frac{1}{\gamma \beta /\Vol}\frac{\text{d}}{\text{d}t}\ln(T) = -V 
\end{align}

Substituting these relations into Eqn.\ \eqref{mfeq} for $\dot{V}$, we obtain
\begin{align}
\dot{V} &= -(p\tau_I/n_I)\left[n_I\left(\dot{T}+\sum_{i=1}^{n_E}\dot{E}_i\right)+\sum_{k=1}^{n_I}\sum_{j=1}^k \dot{I}_j\right]+\frac{c}{\gamma \beta/\Vol}\frac{\text{d}}{\text{d}t} \ln(T)+\dot{T}/\gamma \label{preint}
\end{align}

Integrating Eqn.\ \eqref{preint} from $0$ to $+\infty$ and using the fact that $E_i(0) = E_i(\infty) = I_j(0) = I_j(\infty) = V(\infty) = 0$ for $i=1,2,...,n_E$ and $j=1,2,...,n_I$, it follows that,
\begin{align}
-V(0) = -p\tau_I \left\{T(\infty)-T(0)\right\}+\frac{c}{\gamma \beta /\Vol}\left\{\ln[T(\infty)]-\ln[T(0)]\right\}+(1/\gamma)\left\{T(\infty)-T(0)\right\} \label{V0fsize}
\end{align}

Dividing by $T(0) = N_\text{cells}$ we have
\begin{align}
-V(0)/N_\text{cells} &= (1/\gamma-p\tau_I)[T(\infty)/N_\text{cells}-1]+\frac{c}{\gamma \beta N_\text{cells}/\Vol}\ln(T(\infty)/N_\text{cells}) \nonumber \\
-V(0)/N_\text{cells}+1/\gamma-p\tau_I &= (1/\gamma-p\tau_I)T(\infty)/N_\text{cells}+\frac{c}{\gamma \beta N_\text{cells}/\Vol}\ln(T(\infty)/N_\text{cells}) \nonumber \\
-\gamma V(0)/N_\text{cells}+1-\gamma p\tau_I &= (1-\gamma p\tau_I)T(\infty)/N_\text{cells}+\frac{c}{\beta N_\text{cells}/\Vol}\ln(T(\infty)/N_\text{cells}) \nonumber \\
-\frac{\gamma \beta/\Vol \cdot V(0)}{c} -\frac{(\beta N_\text{cells}/\Vol)\cdot (\gamma p\tau_I-1)}{c} &= -\frac{(\beta N_\text{cells}/\Vol)\cdot (\gamma p\tau_I-1)}{c}T(\infty)/N_\text{cells}+\ln(T(\infty)/N_\text{cells}) \label{fsize1}
\end{align}

Substituting the critical fraction of cells uninfected $\Tstar/N_\text{cells} = c/[(\beta N_\text{cells}/\Vol) \cdot (\gamma p\tau_I-1)]$ (right-hand side of Eqn.\ \eqref{tstar}) into Eqn.\ \eqref{fsize1}, we obtain
\begin{align}
-\frac{\gamma \beta/\Vol \cdot V(0)}{c} -\frac{1}{\Tstar/N_\text{cells}} &= -\frac{T(\infty)/N_\text{cells}}{\Tstar/N_\text{cells}}+\ln(T(\infty)/N_\text{cells}) \nonumber \\
\me^{-[\gamma \beta/\Vol \cdot V(0)]/c} \me^{-1/(\Tstar/N_\text{cells})} &= T(\infty)/N_\text{cells} \cdot \me^{-(T(\infty)/N_\text{cells})/(\Tstar/N_\text{cells})} \nonumber \\
-\frac{\me^{-1/(\Tstar/N_\text{cells})}}{\Tstar/N_\text{cells}}\me^{-[\gamma \beta/\Vol \cdot V(0)]/c}&= -\frac{T(\infty)/N_\text{cells}}{\Tstar/N_\text{cells}} \cdot \me^{-(T(\infty)/N_\text{cells})/(\Tstar/N_\text{cells})} \label{fsize2}
\end{align}

Eqn.\ \eqref{fsize2} is of the form $z = w\exp(w)$. The Lambert $W$ function is the function that gives the inverse relation, i.e.\ $W(z) = w$. Therefore, we have the following,
\begin{align}
-\frac{T(\infty)/N_\text{cells}}{\Tstar/N_\text{cells}} = W_0\hspace{-0.3em}\left(-\frac{\me^{-1/(\Tstar/N_\text{cells})}}{\Tstar/N_\text{cells}}\me^{-[\gamma \beta/\Vol \cdot V(0)]/c}\right) \nonumber \\
T(\infty)/N_\text{cells} = -\Tstar/N_\text{cells} \cdot W_0\hspace{-0.3em}\left(-\frac{\me^{-1/(\Tstar/N_\text{cells})}}{\Tstar/N_\text{cells}}\me^{-[\gamma \beta/\Vol \cdot V(0)]/c}\right) 
\end{align}
where here we make use of $W_0$ the upper branch of the Lambert $W$ function. The upper branch is used because $T(\infty)/\Tstar < 1$ since $T(\infty)$ is always reached after $\Tstar$ and therefore is always smaller.

The fraction of cells consumed by the infection, $1-T(\infty)/N_\text{cells}$, is then given by
\begin{align}
1-T(\infty)/N_\text{cells} = 1+\Tstar/N_\text{cells} \cdot W_0\hspace{-0.3em}\left(-\frac{\me^{-1/(\Tstar/N_\text{cells})}}{\Tstar/N_\text{cells}}\me^{-[\gamma \beta/\Vol \cdot V(0)]/c}\right) \label{fsize3}
\end{align}

If instead there was initially a number of infectious cells but no infectious virions ($I_1(0) \neq 0$, $V(0) = 0$) then Eqn.\ \eqref{V0fsize} would be
\begin{align}
0 = -p\tau_I \left\{T(\infty)-T(0)-I_1(0)\right\}+\frac{c}{\gamma \beta /S}\left\{\ln[T(\infty)]-\ln[T(0)]\right\}+(1/\gamma)\left\{T(\infty)-T(0)\right\} \label{I0fsize}
\end{align}
which is equivalent to Eqn.\ \eqref{V0fsize} where $V(0)$ is replaced with $p \tau_I I_1(0)$. This means that Eqn.\ \eqref{fsize3} would be the same but with $V(0)$ replaced with $p \tau_I I_1(0)$.

For $V(0)$ such that $[\gamma \beta/\Vol \cdot V(0)]/c \approx 0$, Eqn.\ \eqref{fsize3} simplifies to
\begin{align}
1-T(\infty)/N_\text{cells} = 1+\Tstar/N_\text{cells} \cdot W_0\hspace{-0.3em}\left(-\frac{\me^{-1/(\Tstar/N_\text{cells})}}{\Tstar/N_\text{cells}}\right) \label{fsize4}
\end{align}

The Taylor series of $W_0(x)$ around 0 is $W_0(x) = x-x^2+3x^3/2-\dots \approx x$. It follows that, when $\Tstar/N_\text{cells} \ll 1$, Eqn.\ \eqref{fsize4} simplifies to
\begin{align}
1-T(\infty)/N_\text{cells} \approx 1+\Tstar/N_\text{cells} \cdot -\frac{\me^{-1/(\Tstar/N_\text{cells})}}{\Tstar/N_\text{cells}} \nonumber \\
1-T(\infty)/N_\text{cells} \approx 1-\me^{-1/(\Tstar/N_\text{cells})}
\end{align}

\begin{figure} 
\begin{center}
\includegraphics[width=0.4\linewidth]{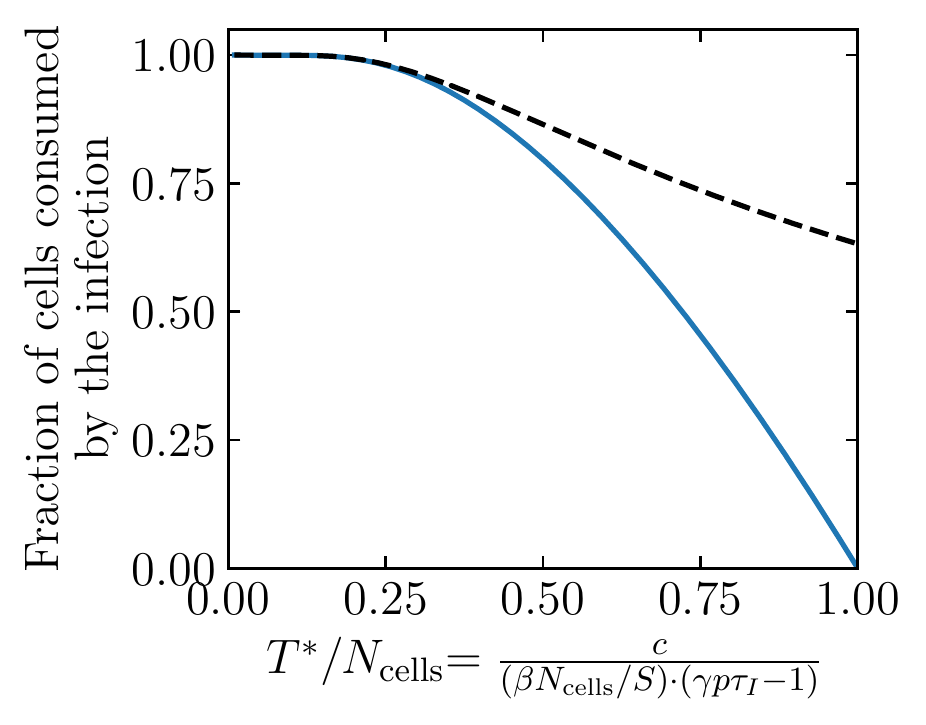}
\caption{\textbf{MFM-predicted fraction of cells consumed by the infection.} The MFM-predicted fraction of cells consumed by the infection ($1-\fsize$, Eqn.\ \eqref{fsize4}, blue solid line) as a function of the fraction of cells uninfected such that the reproductive number is 1 ($\Tstar/N_\text{cells}$). The black dashed line is $1-\exp[-1/(\Tstar/N_\text{cells})]$.}
\label{Tstar}
\end{center}
\end{figure}

Fig~\ref{Tstar} shows that the fraction of cells consumed by the infection, $1-\fsize$, is simply a strictly decreasing function of $\Tstar/N_\text{cells}$ and that, when $\Tstar/N_\text{cells} \ll 1$, $1-\fsize \approx 1-\me^{-1/(\Tstar/N_\text{cells})}$.

Let us now derive an expression for the reproductive number $R(t)$ as a function of the fraction of cells uninfected $T(t)/N_\text{cells}$ for a given basic reproductive number $R_0$  (Eqn.\ \eqref{R0}) and critical fraction of cells uninfected $\Tstar/N_\text{cells}$ (Eqn.\ \eqref{tstar}).

To begin, using Eqn.\ \eqref{R0} and the left-hand side of Eqn.\ \eqref{tstar}, we have the following relation,
\begin{align}
\frac{1}{R_0}\frac{p\tau_I \cdot \gamma \beta N_\text{cells}/\Vol}{c+\beta N_\text{cells}/\Vol} &=\frac{p\tau_I \cdot \gamma \beta \Tstar/\Vol}{c+\beta \Tstar/\Vol} \nonumber \\
\frac{1}{R_0}\frac{N_\text{cells}}{c/(\beta/\Vol)+N_\text{cells}} &=\frac{\Tstar}{c/(\beta/\Vol)+\Tstar} \nonumber \\
N_\text{cells}[c/(\beta/\Vol)]+N_\text{cells}\Tstar &= R_0[c/(\beta/\Vol)]\Tstar+R_0N_\text{cells}\Tstar \nonumber \\
c/(\beta/\Vol)\cdot (N_\text{cells}-R_0\Tstar) &= N_\text{cells}\Tstar(R_0-1) \nonumber \\
c/(\beta/\Vol) &= \frac{N_\text{cells}\Tstar(R_0-1)}{N_\text{cells}-R_0\Tstar} \label{sigma}
\end{align}

The left-hand side of Eqn.\ \eqref{tstar} can be rearranged to
\begin{align}
\gamma p\tau_I &= \frac{c+\beta \Tstar/\Vol}{\beta \Tstar/\Vol} \nonumber \\
\gamma p\tau_I &= \frac{c/(\beta/\Vol)+\Tstar}{\Tstar} \label{gB}
\end{align}

Substituting Eqn.\ \eqref{sigma} into Eqn.\ \eqref{gB}, we obtain 
\begin{align}
\gamma p\tau_I &= \frac{1}{\Tstar}\left[\frac{N_\text{cells}\Tstar(R_0-1)}{N_\text{cells}-R_0\Tstar}+\Tstar\right] \nonumber \\
\gamma p\tau_I &= \frac{N_\text{cells}(R_0-1)}{N_\text{cells}-R_0\Tstar}+1 \nonumber \\
\gamma p\tau_I &= \frac{R_0(N_\text{cells}-\Tstar)}{N_\text{cells}-R_0\Tstar}
\end{align}

The reproductive number can also be written as
\begin{align}
R(t) &= \frac{p\tau_I \cdot \gamma \beta T(t)/\Vol}{c+\beta T(t)/\Vol} \nonumber \\
R(t) &= \frac{\gamma p\tau_I \cdot T(t)}{c/(\beta/\Vol)+T(t)} \label{Ralt}
\end{align}

Substituting Eqn.\ \eqref{sigma} and Eqn.\ \eqref{gB} into Eqn.\ \eqref{Ralt}, we then have
\begin{align}
R(t) &= \frac{\frac{R_0(N_\text{cells}-\Tstar)}{N_\text{cells}-R_0\Tstar} \cdot T(t)}{\frac{N_\text{cells}\Tstar(R_0-1)}{N_\text{cells}-R_0\Tstar}+T(t)} \nonumber \\
R(t) &= \frac{R_0 T(t)\cdot (N_\text{cells}-\Tstar)}{N_\text{cells}\Tstar(R_0-1)+T(t) \cdot (N_\text{cells}-R_0\Tstar)} \nonumber \\
R(t) &= \frac{R_0 \cdot T(t)/N_\text{cells} \cdot (1-\Tstar/N_\text{cells})}{\Tstar/N_\text{cells} \cdot (R_0-1)+T(t)/N_\text{cells} \cdot (1-R_0 \cdot \Tstar/N_\text{cells})} \label{phaseeq}
\end{align}

\subsection{Probability that an infectious virion succeeds at causing a productive cell infection} \label{qsuccess-section}

Let us derive the probability that an infectious virion is successful at causing a productive cell infection, $\qsuccess$, in a population of fully susceptible, uninfected cells ($T=N_\text{cells}$). 

In a time step $\Delta t$, the probability that an infectious virion neither loses infectivity nor enters a cell is $(1-\Delta tc-\Delta t \beta N_\text{cells}/\Vol)$, the probability that an infectious virion enters a cell is $\Delta t\beta_\text{} N_\text{cells}/\Vol$ and the probability that an infectious virion post cell entry will be successful at causing a cell infection is $\gamma$. The probability that an infectious virion is successful at causing a productive cell infection in time $t\,=\,k\Delta t$ is then given by the following expression,
\begin{align}
(1-\Delta tc-\Delta t\beta_\text{} N_\text{cells}/\Vol)^{k-1}(\Delta t \beta_\text{} N_\text{cells}/\Vol)\gamma
\end{align}

More generally, the probability that this happens in any time $t = k\Delta t$ where $0 \leq t \leq t_\text{incub}$ and $t_\text{incub}\,=\,n\Delta t$ is the incubation time, is then expressed as
\begin{align}
&\sum_{k\,=\,1}^n \left[ (1-\Delta tc-\Delta t\beta_\text{} N_\text{cells}/\Vol)^{k-1}(\Delta t \beta_\text{} N_\text{cells}/\Vol) \gamma \right] \nonumber \\
&= \left[ \Delta t \gamma \beta_\text{} N_\text{cells}/\Vol\right]\sum_{k\,=\,1}^n \left[1-\Delta t(c+\beta_\text{} N_\text{cells}/\Vol)\right]^{k-1} \nonumber \\
&= \left[ \Delta t \gamma \beta_\text{} N_\text{cells}/\Vol\right]\underbracket{\sum_{k\,=\,0}^{n-1} \left[1-\Delta t(c+\beta_\text{} N_\text{cells}/\Vol)\right]^k}_{\text{partial sum of geometric series}} \nonumber \\
&= \left[ \Delta t\gamma\beta_\text{} N_\text{cells}/\Vol\right]\left[\frac{1-[1-\Delta t (c+\beta N_\text{cells}/\Vol)]^n}{1-[1-\Delta t (c+\beta N_\text{cells}/\Vol)]}\right] \nonumber \\
&= \left[\Delta t\gamma\beta_\text{} N_\text{cells}/\Vol\right]\left[\frac{1-[1-\Delta t (c+\beta N_\text{cells}/\Vol)]^n}{\Delta t (c+\beta N_\text{cells}/\Vol)}\right] \nonumber \\
&= \left[ \frac{\gamma\beta_\text{} N_\text{cells}/\Vol}{c+\beta N_\text{cells}/\Vol}\right]\left[1-\left(1-\frac{t_\text{incub}(c+\beta N_\text{cells}/\Vol)}{n}\right)^n\right] \label{1fail-prelimit}
\end{align}

Taking the limit of Eqn.\ \eqref{1fail-prelimit} as $n \to \infty$ (i.e.\ $\Delta t \to 0$),
\begin{align}
&\left[ \frac{\gamma\beta_\text{} N_\text{cells}/\Vol}{c+\beta N_\text{cells}/\Vol}\right]\left[1-\lim_{n \to \infty}\left(1-\frac{t_\text{incub}(c+\beta N_\text{cells}/\Vol)}{n}\right)^n\right] \nonumber \\
= &\left[ \frac{\gamma\beta_\text{} N_\text{cells}/\Vol}{c+\beta N_\text{cells}/\Vol}\right]\left[1-\me^{-t_\text{incub}(c+\beta N_\text{cells}/\Vol)}\right] \label{1fail-tincub}
\end{align}

As $t_\text{incub} \to \infty$, Eqn.\ \eqref{1fail-tincub} is then given by the following expression,
\begin{align}
\qsuccess = \frac{\gamma\beta_\text{} N_\text{cells}/\Vol}{c+\beta_\text{} N_\text{cells}/\Vol}  \label{1success}
\end{align}

\begin{figure}
\begin{center}
\includegraphics[width=\linewidth]{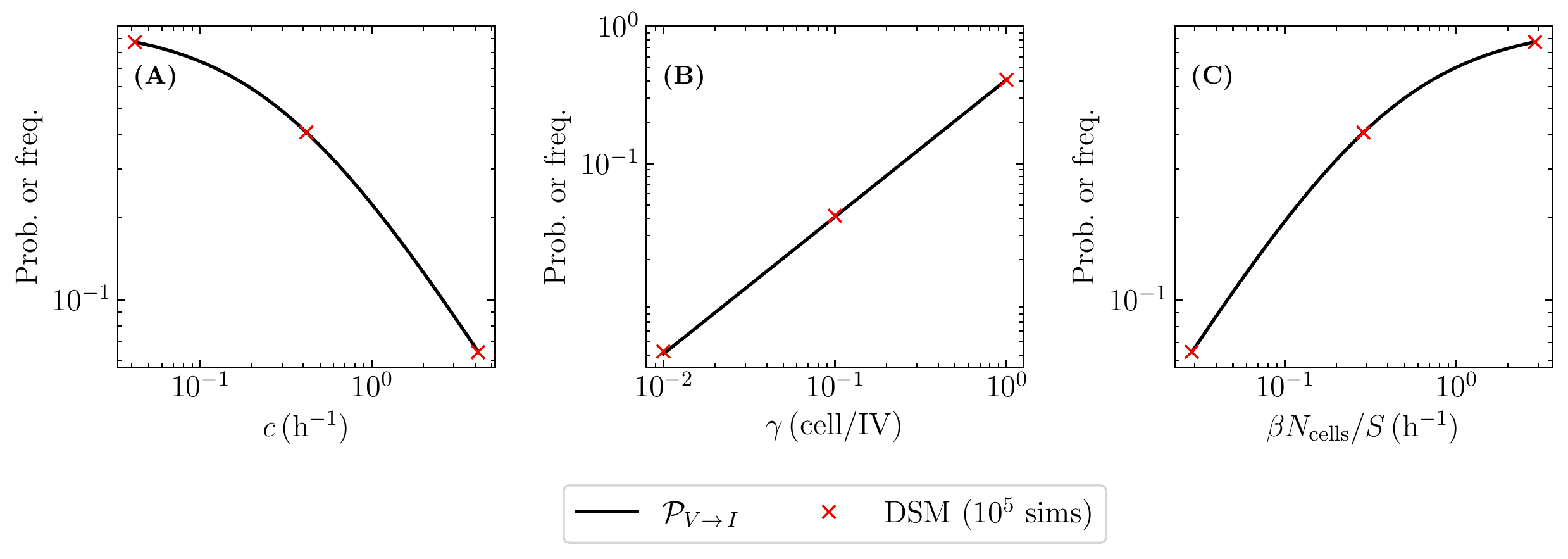}
\end{center}
\caption{\textbf{Probability that an infectious virion is successful at causing a productive cell infection.} Frequency of success of an infectious virion to cause a productive cell infection in a population of uninfected cells for $10^5$ DSM simulations (red scatter) compared to $\qsuccess$ the derived expression for the probability of success of an infectious virion to cause a productive cell infection in a population of uninfected cells (black curve, Eqn.\ \eqref{1success}) while either varying (A) $c$, (B) $\gamma$ or (C) $\beta_\text{}N_\text{cells}/\Vol$. Unless otherwise specified, the parameters were the same as in Fig~\ref{mrhoV}.}
\label{1fail-fig}
\end{figure}

Fig~\ref{1fail-fig} shows that the frequency of success of an infectious virion to cause a productive cell infection in a population of uninfected cells, generated from $10^5$ DSM simulations, is in agreement with Eqn.\ \eqref{1success}, over a wide range of infection parameters.

\subsection{Probability that an infectious cell produces $m$ infectious virions} \label{qprod-section}

Let us derive the probability that an infectious cell produces $m$ infectious virions over its lifespan ($\qprod$).

Assuming that successive infectious virion productions are independent events that happen under constant rate $p$, then the probability that an infectious cell produces $m$ infectious virions over its lifespan $t$ is given by $\text{Poisson}(m|\lambda = t\,p)$.

The negative binomial (NB) distribution represents the probability that there are $k$ successes in a sequence of independent and identically distributed Bernoulli trials with probability of success $p$ before a specified number of failures $r$ occurs, namely,
\begin{align} 
\text{NB}(k|r,\,p) = {k+r-1 \choose k}(1-p)^rp^m
\end{align}

The probability that an infectious cell transition to the next compartment is given by $\text{Binomial}(1|n=1,\, p_I= \Delta tn_I/\tau_I)$. Therefore the probability that an infectious cell transition to the next compartment for $n_I$ time steps and does not for $n-n_I$ time steps is then given by $\text{NB}(n_I|r=n-n_I,\, p_I=\Delta t\, n_I/\tau_I)$. It is possible to show that the continuous analogue of this discrete probability distribution is the Erlang distribution ($\text{Erlang}(t|k=n_I,\, \lambda = n_I/\tau_I)$). As a result, we make use of this probability density function to express the probability that the infectious lifespan has duration $t$.

The probability that an infectious cell produces $m$ infectious virions over its lifespan, i.e.\ that the burst size of an infected cell is $m$, can then be derived by integrating over time $t$, the joint probability distribution that an infectious cell produces $m$ infectious virions given a lifespan of duration $t$ and that an infectious cell has a lifespan of duration $t$, namely,
\begin{align}
\qprod = \int_0^\infty \text{Poisson}(m|\lambda=pt) \cdot \text{Erlang}(t|k=n_I,\, \lambda=n_I/\tau_I) \text{d}t = \text{NB}(m|r=n_I,\, p_B=\B/(n_I+\B))  \label{burst-size}
\end{align}
where $\B = p\tau_I$ is the average burst size. When $n_I = 1$, $\qprod = \text{NB}(m|r=1,\, p_B=\B/(1+\B)) = \text{Geom}(m|p=1/(1+\B))$ where Geom represents the geometric distribution. When $n_I \to \infty$, $\qprod = \text{NB}(m|r=n_I,\, p_B=\B/(n_I+\B)) = \text{Poisson}(m|\lambda = \B)$.

\begin{figure}
\begin{center}
\includegraphics[width=\linewidth]{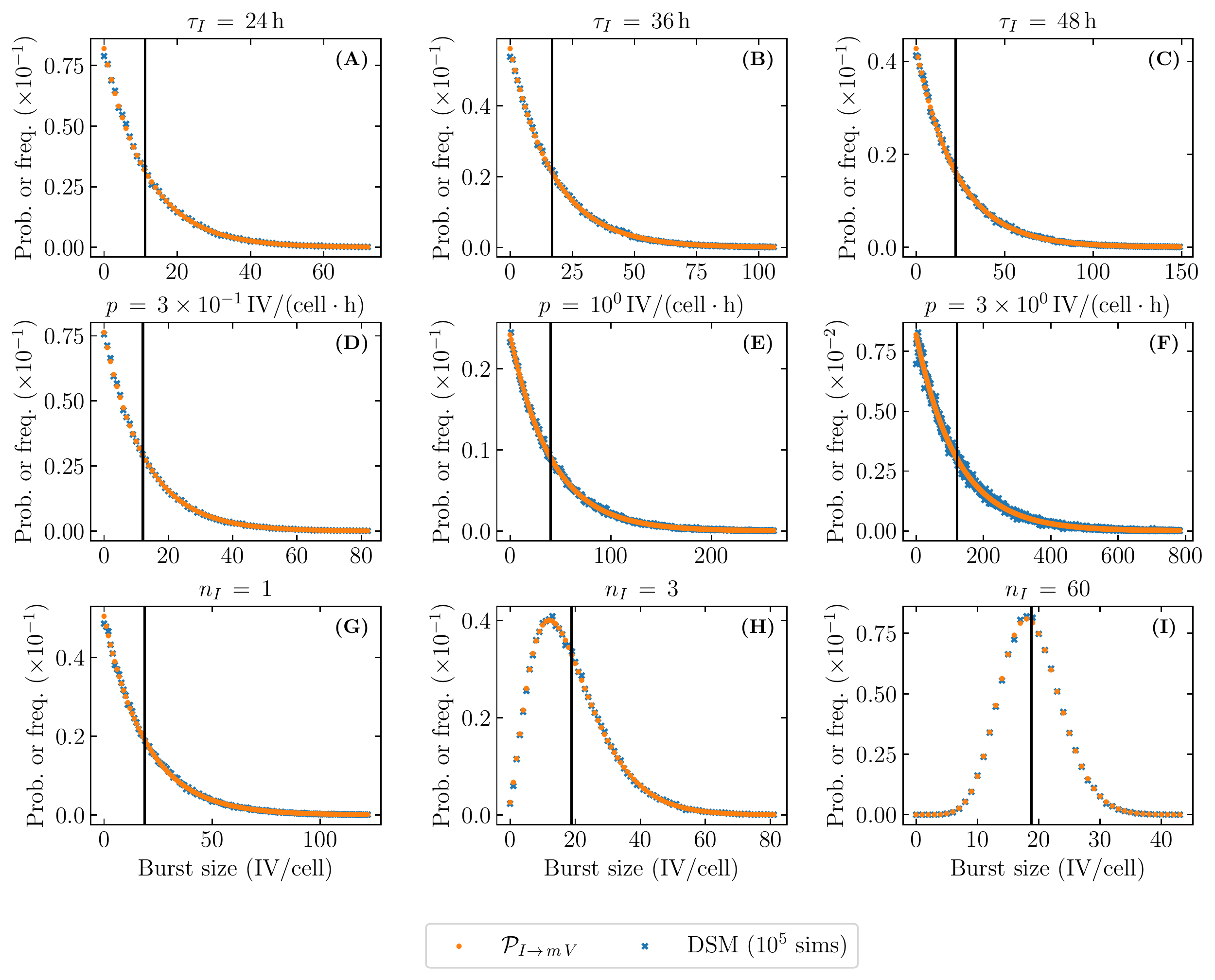}
\end{center}
\caption{\textbf{Probability that an infectious cell will produce $m$ infectious virions.} Normalized histogram of the burst size, i.e.\ the total number of infectious virions produced by one infectious cell over its lifespan, for $10^5$ DSM simulations (blue x), compared to the derived expression for the probability mass function of the burst size distribution (orange circle, Eqn.\ \eqref{burst-size}) while either varying (A--C) $\tau_I$, (D--F) $p$ or (G--I) $n_I$. The black lines show the average burst size values $\B = p \tau_I$. Unless otherwise specified, the parameters were the same as in Fig~\ref{mrhoV}.}
\label{burst-size-fig}
\end{figure}

Fig~\ref{burst-size-fig} shows that the normalized histogram of the burst size, generated from $10^5$ DSM simulations, is in agreement with Eqn.\ \eqref{burst-size}, over a wide range of infection parameters.

\subsection{Extinction probability} \label{rhoV-section}

Let us derive the extinction probability of an infection. Following \pearson, the extinction probability of an infection with initially one infectious virion ($\rhoV$) can be written as
\begin{align}
\rhoV = \qfail+\qsuccess \cdot \rhoI \label{failV}
\end{align}
since the infection can become extinct if the initial infectious virion fails to cause a productive cell infection with probability $\qfail$ or if it does cause a productive cell infection with probability $\qsuccess=(1-\qfail)$ but that cell infection then leads to extinction with probability $\rhoI$. 

Likewise, the extinction probability of an infection with initially one infectious cell ($\rhoI$) can be expressed as
\begin{align}
\rhoI = \sum_{m=0}^\infty \qprod \cdot \left(\rhoV\right)^m \label{rhoI}
\end{align}
since the initial infectious cell will produce some number $m$ of infectious virions over its lifespan with probability $\qprod$ and each one of these produced infectious virions can be seen as an independent infection event that can lead to extinction with probability $\rhoV$.

Substituting our derived expression for $\qprod$ (Eqn.\ \eqref{burst-size}) into Eqn.\ \eqref{rhoI} yields
\begin{align}
\rhoI &= \sum_{m=0}^\infty \text{NB}(m|r=n_I,p_B=\B/(n_I+\B)) \cdot \left(\rhoV\right)^m \nonumber \\
\rhoI &= \sum_{m=0}^\infty {m+n_I-1 \choose m}\left[1-\frac{\B}{n_I+\B}\right]^{n_I}\left[\frac{\B}{n_I+\B}\right]^m \cdot \left(\rhoV\right)^m \nonumber \\
\rhoI &= \left[\frac{n_I}{n_I+\B}\right]^{n_I} \sum_{m=0}^\infty {m+n_I-1 \choose m}\left[\frac{\B}{n_I+\B}\cdot \rhoV\right]^m \label{failIstep1}
\end{align}

The following binomial series can be used to simplify Eqn.\ \eqref{failIstep1},
\begin{align}
\frac{1}{(1-x)^s} = \sum_k {s+k-1\choose k}x^k 
\end{align}

Let us derive some related expressions,
\begin{align}
x &\coloneqq \frac{\B}{n_I+\B}\cdot \rhoV \\
\Longrightarrow 1-x &= \frac{n_I+\B(1-\rhoV)}{n_I+\B} \nonumber \\
\Longrightarrow \frac{1}{(1-x)^{n_I}} &= \left[ \frac{n_I+\B}{n_I+\B(1-\rhoV)} \right]^{n_I} 
\end{align}

Eqn.\ \eqref{failIstep1} can then be simplified to
\begin{align}
 \rhoI &= \left[\frac{n_I}{n_I+\B}\right]^{n_I} \left[\frac{n_I+\B}{n_I+\B(1-\rhoV)}\right]^{n_I} \nonumber \\
 &= \left[\frac{n_I}{n_I+\B(1-\rhoV)}\right]^{n_I} \nonumber \\
 &= \left[\frac{\B(1-\rhoV)}{n_I}+1\right]^{-n_I} \label{failIstep2}
\end{align}

Substituting Eqn.\ \eqref{failIstep2}, our derived expression for $\qsuccess$ (Eqn.\ \eqref{1success}) and $\qfail = 1-\qsuccess$ into Eqn.\ \eqref{failV} yields
\begin{align}
\rhoV &= \qfail+\qsuccess \left[\frac{\B(1-\rhoV)}{n_I}+1\right]^{-n_I} \label{preroot} \\
\rhoV &= \left[1-\frac{\gamma\beta_\text{} N_\text{cells}/\Vol}{c+\beta_\text{} N_\text{cells}/\Vol}\right]+\frac{\gamma\beta_\text{} N_\text{cells}/\Vol}{c+\beta_\text{} N_\text{cells}/\Vol} \left[\frac{\B(1-\rhoV)}{n_I}+1\right]^{-n_I} \nonumber \\
\rhoV &= 1-\frac{\gamma}{c/(\beta N_\text{cells}/\Vol)+1}+\frac{\gamma}{c/(\beta N_\text{cells}/\Vol)+1}\left[\frac{\B(1-\rhoV)}{n_I}+1\right]^{-n_I} \nonumber \\
0 &= 1-\frac{[1+c/(\beta N_\text{cells}/\Vol)]}{\gamma}[1-\rhoV]-\left[\frac{\B(1-\rhoV)}{n_I}+1\right]^{-n_I} \label{root}
\end{align}

\begin{figure}
\begin{center}
\includegraphics[width=\linewidth]{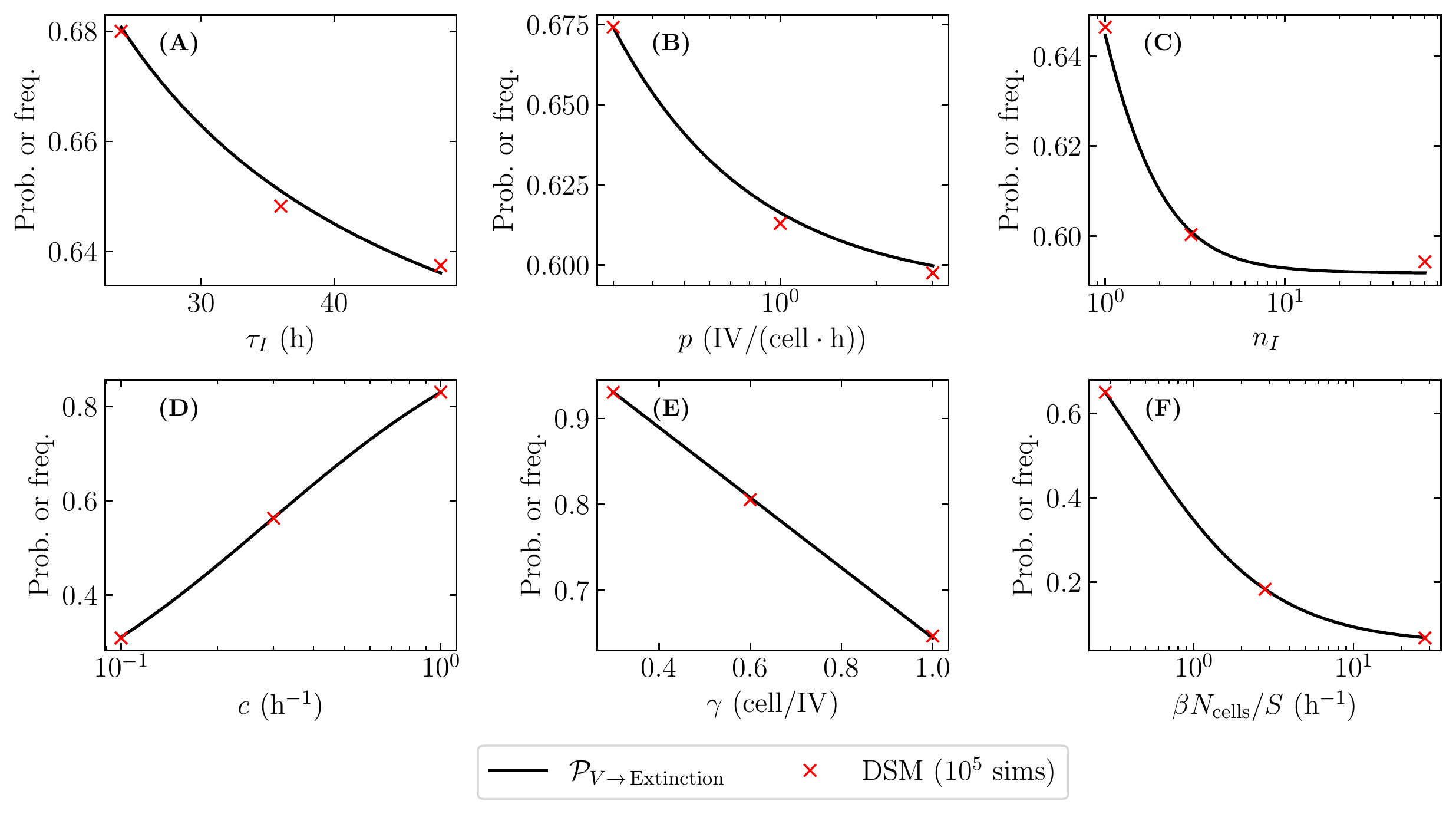}
\end{center}
\caption{\textbf{Extinction probability given an infection initiated with only one infectious virion.} Frequency of failure of an infection initiated with only one infectious virion to cause more than 0.01\% of cells to be infected by the end for $10^5$ DSM simulations (red scatter) compared to $\rhoV$ the derived expression for the extinction probability given an infection initiated with only one infectious virion (black curve, Eqn.\ \eqref{root}) while either varying (A) $\tau_I$, (B) $p$, (C) $n_I$, (D) $c$, (E) $\gamma$ or (F) $\beta N_\text{cells}/\Vol$. Unless otherwise specified, the parameters were the same as in Fig~\ref{mrhoV}.}
\label{rhoV-fig}
\end{figure}

This expression does not seem to have an analytical solution for $\rhoV$ but a solution can be found numerically by finding the roots of the expression on the right-hand side of Eqn.\ \eqref{root} using \texttt{scipy.optimize.fsolve}. Fig~\ref{rhoV-fig} shows that the frequency of extinction given an infection initiated with only one infectious virion, generated from $10^5$ DSM simulations, is in agreement with the probability of extinction given an infection initiated with only one infectious virion, over a wide range of infection parameters. The extinction probability given any number of infectious virions $V_0$ and infectious cells $I_0$ is given by $\left(\rhoV\right)^{V_0}\cdot \left(\rhoI\right)^{I_0}$. It is important to note that, as others \cite{pearson11,yan16}, we also make the assumption in this derivation that $T = N_\text{cells}$ is constant.

When $n_I = 1$, Eqn.\ \eqref{preroot} simplifies to
\begin{align}
\rhoV &= \qfail+\qsuccess \left[\frac{1}{1+\B(1-\rhoV)}\right] 
\end{align}

In this case, $\rhoV = \qfail+1/\B$ is a solution to the above expression as,
\begin{align}
\qfail+1/\B &= \qfail+\qsuccess \left[\frac{1}{1+\B[1-(\qfail+1/\B)]}\right] \nonumber \\
\qfail+1/\B &= \qfail+\qsuccess \left[\frac{1}{\B(1-\qfail)}\right] \nonumber \\
\qfail+1/\B &= \qfail+\qsuccess \left[\frac{1}{\B\cdot\qsuccess}\right] \nonumber \\
\qfail+1/\B &= \qfail+1/\B \nonumber \\
\therefore \text{LHS} &= \text{RHS} \nonumber
\end{align}

$\rhoV = \qfail+1/\B$ is equivalent to the expression in \cite{conway18} (Eq.\ (6) therein). $\rhoV = \qfail+1/\B$ can also be written in terms of the basic reproductive number $R_0 = \B \cdot \qsuccess$ (see Eqn.\ \eqref{R0}), i.e.\  $\rhoV = 1-(R_0-1)/\B$. Also, $\mrhoV$ or  $1 - \rhoV$ is then given by $\qsuccess-1/\B$ or $(R_0-1)/\B$.

When $n_I \to \infty$, $\qprod = \text{Poisson}(m|\lambda = \B)$, therefore, we obtain,
\begin{align}
\rhoV = \qfail+\qsuccess \cdot \sum_{m=0}^\infty \text{Poisson}(m|\lambda=\B) \cdot \left(\rhoV\right)^m 
\end{align}
which is equivalent to the expression in \pearson for their random burst production [DSM] (Eq.\ (26) therein).

%%%%%%%%%%%%%%%%%%%%%%%%%%%%%%%%%%%%%%%%%%%%%%%%%%%%%%%%%%%%%%%%%%%%%%%%%%%%%%%%
%%%%%%%%%%%%%%%%%%%%%%%%%%%%%%%%%%%%%%%%%%%%%%%%%%%%%%%%%%%%%%%%%%%%%%%%%%%%%%%%
%%%%%%%%%%%%%%%%%%%%%%%%%%%%%%%%%%%%%%%%%%%%%%%%%%%%%%%%%%%%%%%%%%%%%%%%%%%%%%%%
%%%%%%%%%%%%%%%%%%%%%%%%%%%%%%%%%%%%%%%%%%%%%%%%%%%%%%%%%%%%%%%%%%%%%%%%%%%%%%%%

\clearpage
\section{Acknowledgements}

This work was supported in part by Discovery Grants 355837-2013 and 2022-03744 (CAAB) from the Natural Sciences and Engineering Research Council of Canada (\url{www.nserc-crsng.gc.ca}) and by the Kato Sechi Female PI Incentive (\url{www.riken.jp/en/careers/programs/kato_sechi}) and Interdisciplinary Theoretical and Mathematical Sciences (iTHEMS, \url{ithems.riken.jp}) programmes at RIKEN (CAAB).

\clearpage
\addcontentsline{toc}{section}{References}
\bibliographystyle{abbrvurl}
\bibliography{flu-stochastic-1}

%%%%%%%%%%%%%%%%%%%%%%%%%%%%%%%%%%%%%%%%%%%%%%%%%%%%%%%%%%%%%%%%%%%%%%%%%%%%%%%%
%%%%%%%%%%%%%%%%%%%%%%%%%%%%%%%%%%%%%%%%%%%%%%%%%%%%%%%%%%%%%%%%%%%%%%%%%%%%%%%%
%%%%%%%%%%%%%%%%%%%%%%%%%%%%%%%%%%%%%%%%%%%%%%%%%%%%%%%%%%%%%%%%%%%%%%%%%%%%%%%%
%%%%%%%%%%%%%%%%%%%%%%%%%%%%%%%%%%%%%%%%%%%%%%%%%%%%%%%%%%%%%%%%%%%%%%%%%%%%%%%%

\clearpage
\begin{appendix}
\renewcommand{\thesection}{S\arabic{section}}
\renewcommand{\theequation}{S\arabic{equation}}
\renewcommand{\thefigure}{S\arabic{figure}}
\renewcommand{\thetable}{S\arabic{table}}
\setcounter{equation}{0}
\setcounter{figure}{0}
\setcounter{table}{0}
\setcounter{page}{1}
%\makeatletter

%\title{\textbf{Supplementary Material for} \protect\\ Stochastic failure of cell infection post viral entry: Implications for infection outcomes and antiviral therapy}
\title{\textbf{Supplementary Material for}\\ \mainmstitle}
\maketitle

\section{Inoculum size} \label{V0-section}

\begin{figure}[h] 
\begin{center}
\includegraphics[width=0.7\linewidth]{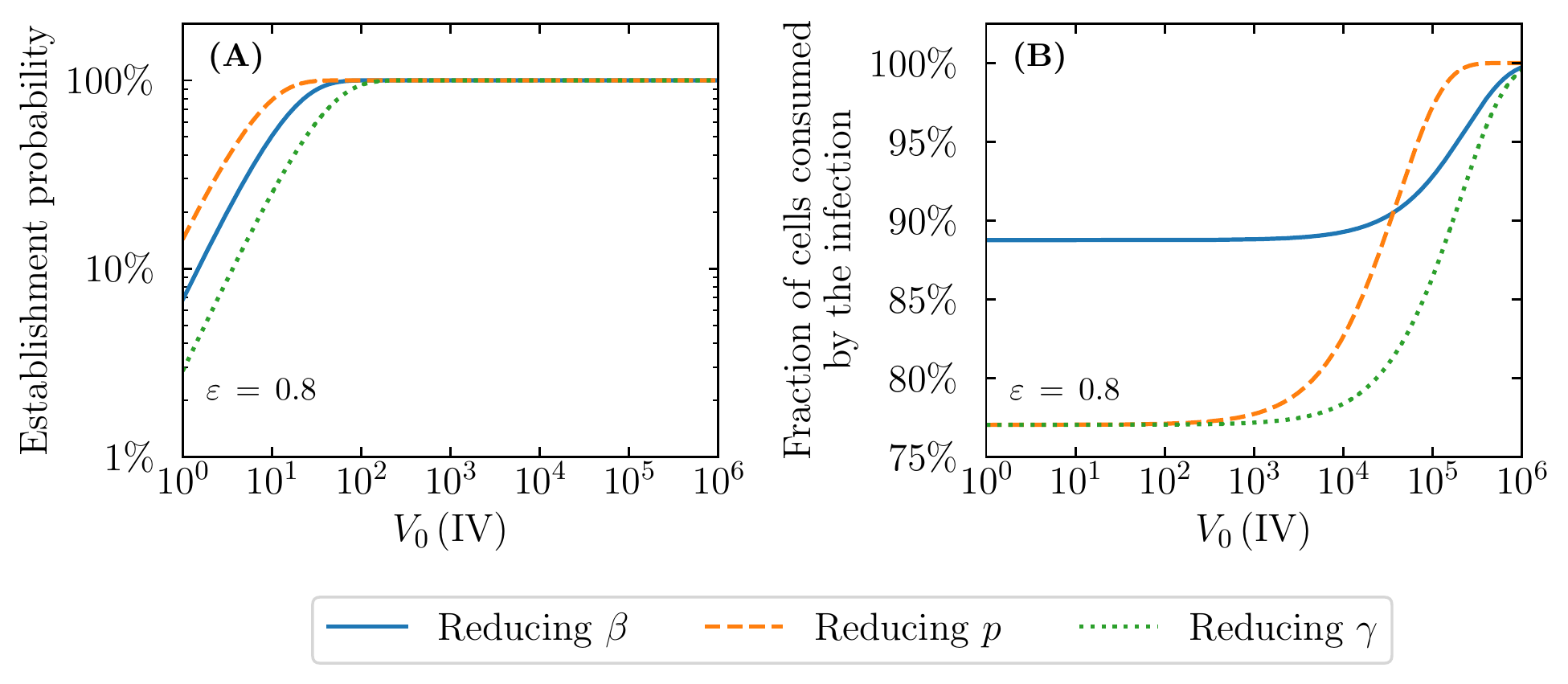}
\caption{\textbf{Inoculum size.} (A) The establishment probability or (B) the MFM-predicted fraction of cells consumed by the infection as a function of the initial number of infectious virions $V_0$ for antivirals with efficacy $\varepsilon = 0.8$ acting either to reduce the virus entry rate, $\beta \to (1-\varepsilon)\beta$ (blue solid line), the virus production rate, $p \to (1-\varepsilon)p$ (orange dashed line), or the probability of a successful cell infection post viral entry, $\gamma \to (1-\varepsilon)\gamma$ (green dotted line). The parameters were the same as in Fig~\ref{mrhoV}.}
\label{V0}
\end{center}
\end{figure}

In the main text, the initial number of virions $V_0 = \unit{1}{\IU}$ and we explored varying the efficacy $\varepsilon$. Now, we fix $\varepsilon = 0.8$ and we explore varying $V_0$. Fig~\ref{V0}(A) and Fig~\ref{V0}(B) shows the establishment probability or the MFM-predicted fraction of cells consumed by the infection respectively as a function of $V_0$ for antivirals with efficacy $\varepsilon = 0.8$ acting on $\beta$, $p$ or $\gamma$. As $V_0$ increases, the establishment probability tends to 100\% and differences in the establishment probability for the 3 antiviral modes of action disappear. For $V_0 \ll N_\text{cells} = \unit{4 \times 10^4}{cells}$, the MFM-predicted fraction of cells consumed by the infection is insensitive to $V_0$. This suggests that this would also be true for the median of the DSM simulated distribution of the fraction of cells consumed by established infections.

\section{Parameter set with a higher burst size} \label{highN-section}

\begin{table*}[h]
\begin{center}
\caption{Parameter sets used in \czuppon.}
\label{modparams}
\begin{tabular}{cllllll}
Parameter set & $p$ [\IU/($\text{cell} \cdot \text{h})$] & $N_\text{cells}$ [cells] & ${R_0}^*$ & $\tau_I$ [h] & $c$ [h$^{-1}$] & $n_I$ \\
\hline
Lower burst size & 11.2/24 & $4 \times 10^4$ & 7.69 & 24/0.595 & 10/24 & 1 \\
Higher burst size & 112/24 & $4 \times 10^3$ & 7.69 & 24/0.595 & 10/24 & 1 \\
\hline
\end{tabular}\\[0.5em]
\begin{minipage}{0.7\linewidth}
* $\beta = [cR_0/\tau_I]/[N_\text{cells}(p-R_0/\tau_I)]$ 
\end{minipage}
\end{center}
\end{table*}

Here, we investigate the other parameter set explored in \czuppon. The difference between the two parameter sets is a 10-fold decrease of the number of cells $N_\text{cells}$, a 10-fold increase of the virus production rate $p$ (hence, of the average burst size $\B = p\tau_I$) and a corresponding $\sim$10-fold decrease of the virus entry rate $\beta N_\text{cells}/\Vol$ (see Table \ref{modparams}). 

\begin{figure}
\begin{center}
\includegraphics[width=0.84\linewidth]{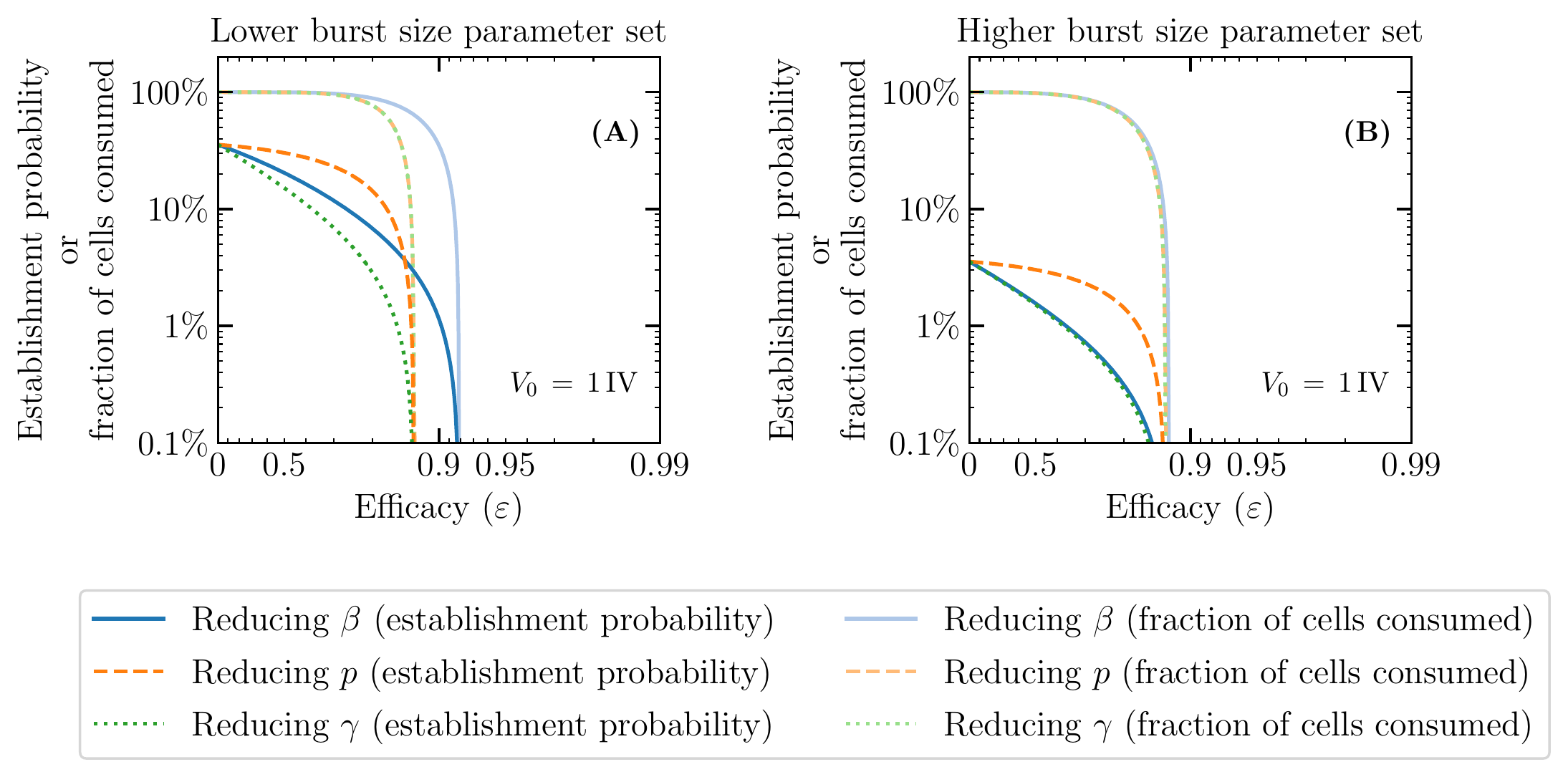}
\caption{\textbf{Lower vs.\ higher burst size parameter set.} The establishment probability (dark colours) or the MFM-predicted fraction of cells consumed by the infection (pale colours) given that there is initially only one infectious virion as a function of efficacy ($\varepsilon$) for antivirals acting either to reduce the virus entry rate, $\beta \to (1-\varepsilon)\beta$ (blue solid lines), the virus production rate, $p \to (1-\varepsilon)p$ (orange dashed lines), or the probability of a successful cell infection post viral entry, $\gamma \to (1-\varepsilon)\gamma$ (green dotted lines) for the parameter set with (A) the lower burst size or (B) the higher burst size (see Table \ref{modparams}) where our additional infection parameters were set to $\gamma = \unit{1}{cell/\IU}$ and $S = \unit{1}{mL}$.}
\label{highN}
\end{center}	
\end{figure}

Fig~\ref{highN} shows both the establishment probability and the MFM-predicted fraction of cells consumed by the infection given that there is initially only one infectious virion as a function of antiviral efficacy ($\varepsilon$) for antivirals acting on $\beta$, $p$ or $\gamma$ for the parameter set with (A) the lower burst size ($\B = \unit{18.8}{\IU/cell}$) or (B) the higher burst size ($\B = \unit{188}{\IU/cell}$). 

For the parameter set with the higher burst size, an antiviral acting on $\gamma$ or $\beta$ have a similar effect on the establishment probability, better than that for an antiviral acting on $p$. Due to the higher burst size, the establishment probability is approximately given by the probability that the initial infectious virion causes a productive cell infection, $\mrhoV = \qsuccess - 1/\B \approx \qsuccess$ for $n_I = 1$ (see Methods, Section \ref{rhoV-section} for more details). The probability that the initial infectious virion will cause a productive cell infection is given by the ratio between the rate of successful cell infection per infectious virion and the rate of virion loss $(\gamma \beta N_\text{cells}/\Vol)/(c + \beta N_\text{cells}/\Vol)$. Since the rate of virion entry into cells is much lower than the rate of virion loss of infectivity ($\beta N_\text{cells}/\Vol \ll c$) the rate of virion loss is mostly governed by the rate of virion loss of infectivity ($(c + \beta N_\text{cells}/\Vol) \approx c$). Therefore, the establishment probability is approximately given by $(\gamma \beta N_\text{cells}/\Vol)/c$ which is affected by $\gamma$ or $\beta$ the same but not affected by $p$. 

For the parameter set with the higher burst size, all antivirals affect comparably the MFM-predicted fraction of cells consumed by the infection. The MFM-predicted fraction of cells consumed is a function of the fraction of cells uninfected such that the reproductive number is 1, i.e.\ $\Tstar/N_\text{cells} = c/[(\beta N_\text{cells}/\Vol) \cdot (\gamma p\tau_I-1)]$ (see Methods, Section \ref{final-size-section} for more details). Due to the higher burst size ($\B = p\tau_I$), $\gamma p \tau_I-1 \approx \gamma p \tau_I$. Therefore, the MFM-predicted fraction of cells consumed is approximately a function of $c/[(\beta N_\text{cells}/\Vol) \cdot (\gamma p\tau_I)]$ which is affected by $\beta$, $p$ or $\gamma$ the same. This suggests that all antivirals also have a similar effect on the median of the DSM simulated distribution of the fraction of cells consumed by established infections.

\section{Post-exposure antiviral therapy} \label{PET-section}

In the main text, we have considered pre-exposure antiviral therapy for an infection initiated with a number of infectious virions. Now, as \czuppon have done, let us also explore antiviral therapy for an infection initiated with only one infectious cell. This may be representative of post-exposure antiviral therapy as it is possible that by the time an antiviral has been given after exposure, the virus has had time to cause some infectious cells.

With $n_I = 1$, the establishment probability given that there is initially one infectious cell (see Methods, Section \ref{rhoV-section} for derivation) is given by
\begin{align}
\mrhoI = 1-\frac{1}{R_0} = 1-\frac{c+\beta N_\text{cells}/\Vol}{p\tau_I \cdot \gamma \beta N_\text{cells}/\Vol} \label{mrhoI}
\end{align}

\begin{figure}[h]
\begin{center}
\includegraphics[width=0.84\linewidth]{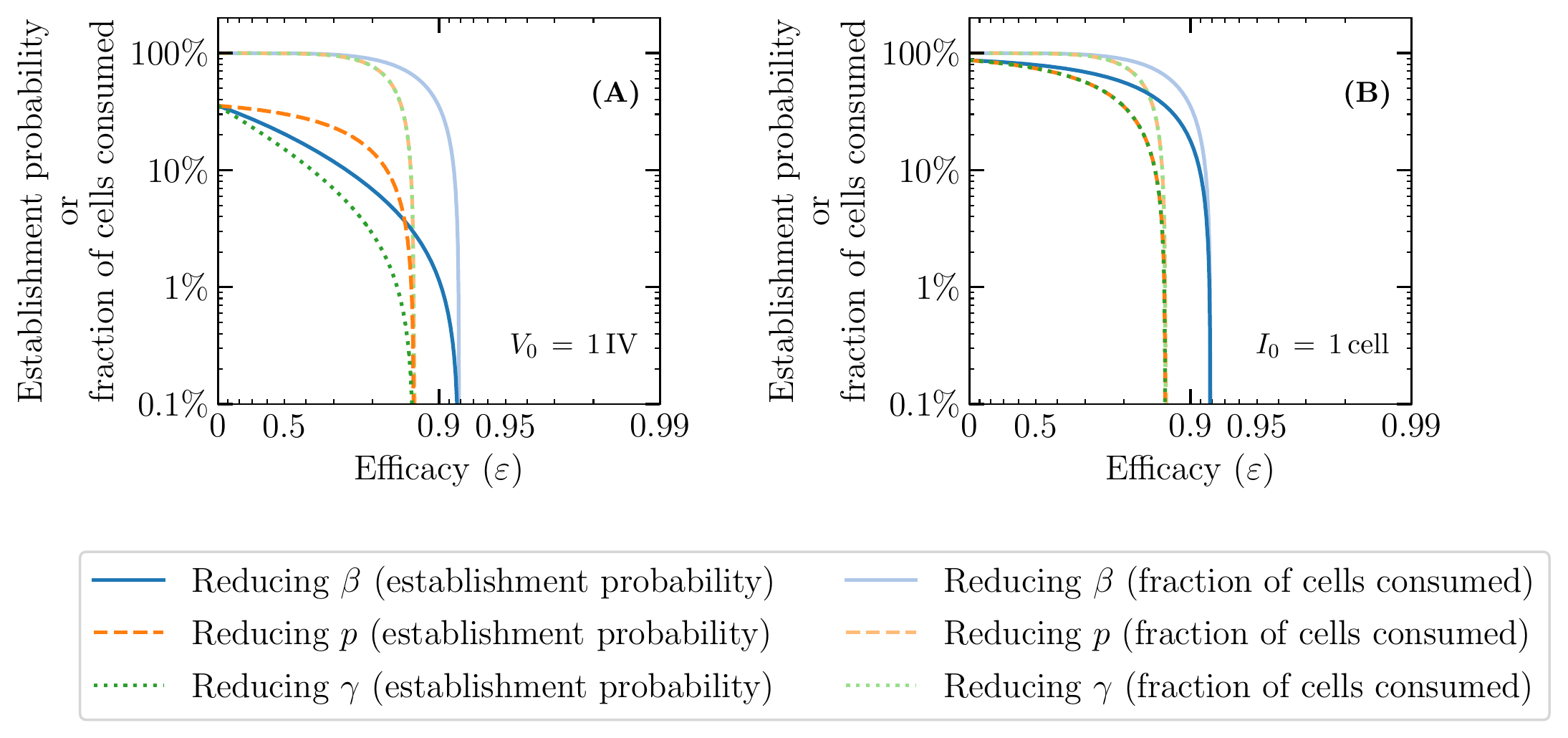}
\caption{\textbf{Pre-exposure vs.\ post-exposure antiviral therapy.} The establishment probability (dark colours) or the MFM-predicted fraction of cells consumed by the infection (pale colours) given that there is initially only (A) one infectious virion or (B) one infectious cell as a function of efficacy ($\varepsilon$) for antivirals acting either to reduce the virus entry rate, $\beta \to (1-\varepsilon)\beta$ (blue solid lines), the virus production rate, $p \to (1-\varepsilon)p$ (orange dashed lines), or the probability of a successful cell infection post viral entry, $\gamma \to (1-\varepsilon)\gamma$ (green dotted lines). The parameters were the same as in Fig~\ref{mrhoV}.}
\label{I0}
\end{center}	
\end{figure}

Fig~\ref{I0} shows the establishment probability and the MFM-predicted fraction of cells consumed by the infection given that there is initially only (A) one infectious virion or (B) one infectious cell as a function of antiviral efficacy ($\varepsilon$) for antivirals acting on $\beta$, $p$ or $\gamma$.

Like \czuppon, with initially one infectious cell, we find that an antiviral acting on $p$ is better than an antiviral acting on $\beta$ to reduce the establishment probability. Reducing $p$ affects only the denominator in the expression for the establishment probability given that there is initially one infectious cell (Eqn.\ \eqref{mrhoI}). Whereas, reducing $\beta$ affects both the numerator and denominator (Eqn.\ \eqref{mrhoI}). Unlike \czuppon, with initially one infectious cell, we find that an antiviral acting on $p$ or $\gamma$ have the same effect on the establishment probability (Eqn.\ \eqref{mrhoI}). 

In addition, there is no noticeable difference in the MFM-predicted fraction of cells consumed by the infection given that there is initially one infectious virion or one infectious cell. The initial number of infectious virions or cells is small and hence has a negligible effect on the MFM-predicted fraction of cells consumed by the infection (see Eqn.\ \eqref{fsize4}). This suggests that this would also be true for the median of the DSM simulated distribution of the fraction of cells consumed by established infections.

\clearpage

\section{Additional figures}

\begin{figure}[h]
\begin{center}
\includegraphics[width=0.7\linewidth]{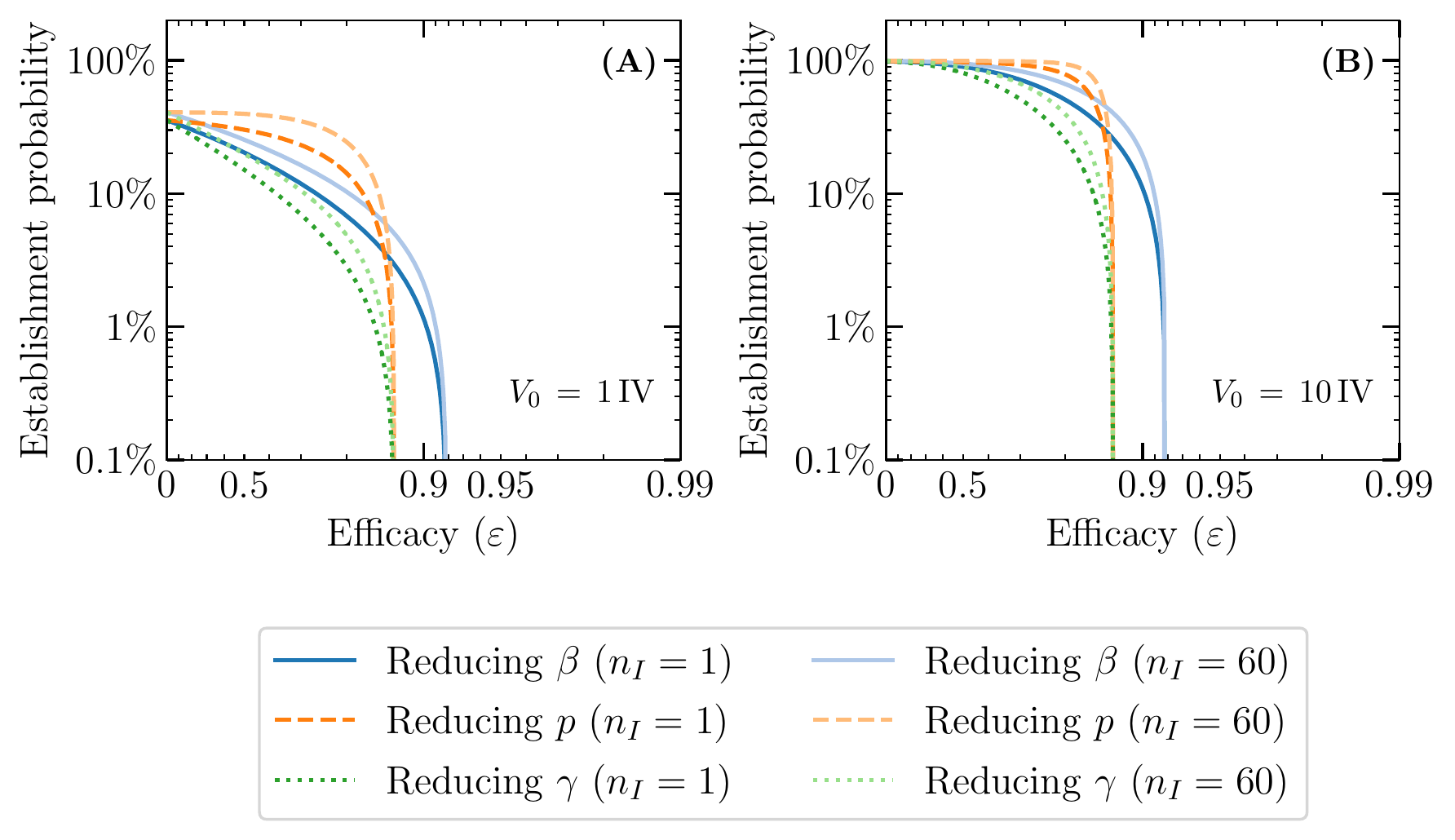}
\caption{\textbf{Effect of infectious phase duration distribution on the ability of antivirals to reduce the establishment probability.} The establishment probability given an infection initiated with (A) 1 infectious virion or (B) 10 infectious virions as a function of efficacy ($\varepsilon$) for antivirals acting either to reduce the virus entry rate, $\beta \to (1-\varepsilon)\beta$ (blue solid lines), the virus production rate, $p \to (1-\varepsilon)p$ (orange dashed lines), or the probability of a successful cell infection post viral entry, $\gamma \to (1-\varepsilon)\gamma$ (green dotted lines) where either $n_I = 1$ (dark colours) or $n_I = 60$ (pale colours). Unless otherwise specified, the parameters were the same as in Fig~\ref{mrhoV}.}
\label{mrhoVnI}
\end{center}
\end{figure}

\begin{figure}[h]
\begin{center}
\includegraphics[width=1.0\linewidth]{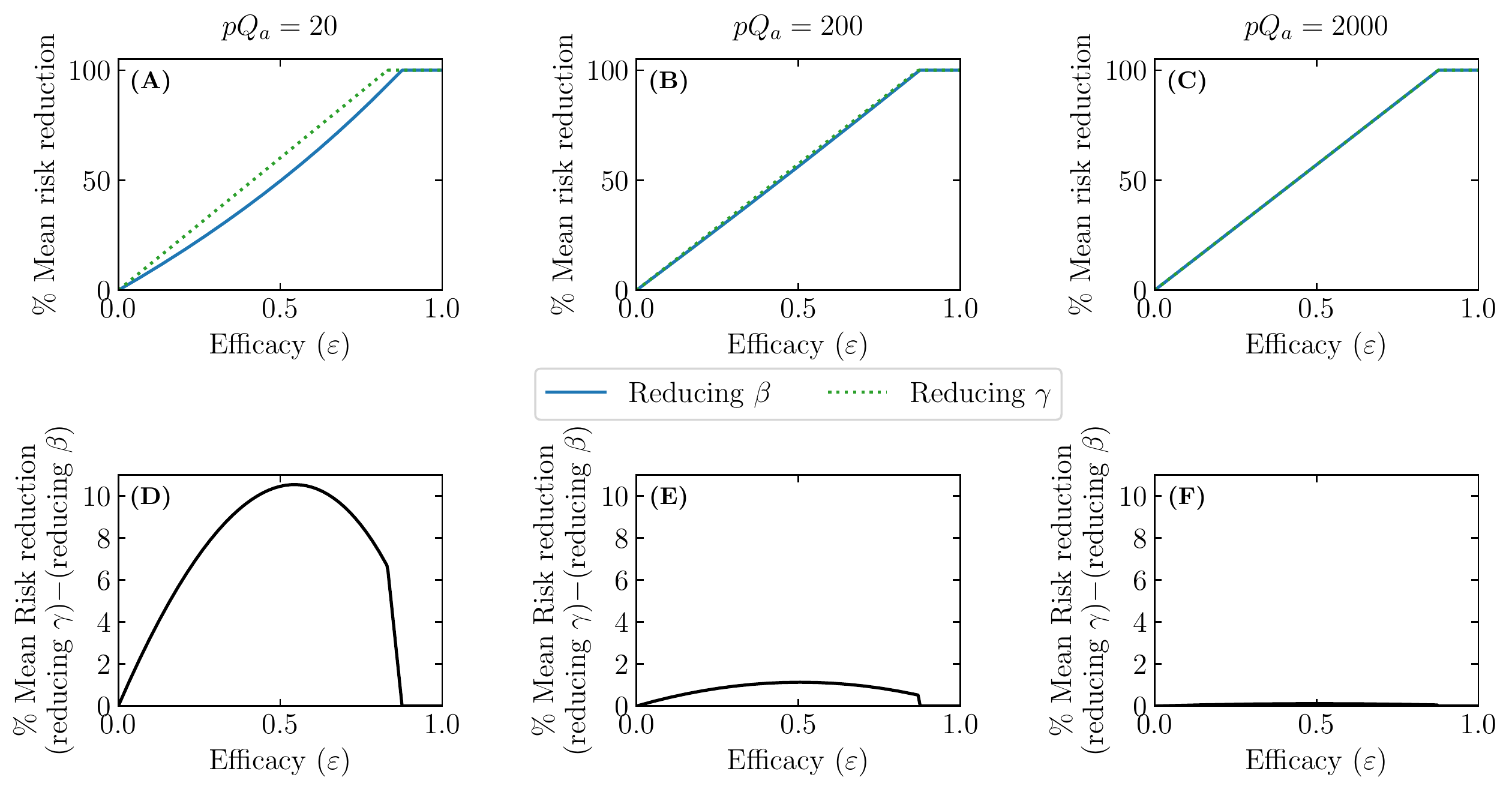}
\caption{\textbf{Mean infection risk reduction.} (A--C) Mean risk reduction as a function of efficacy ($\varepsilon$) for antivirals acting either to reduce the virus entry rate, $\beta \to (1-\varepsilon)\beta$ (blue solid line), or the probability of a successful cell infection post viral entry, $\gamma \to (1-\varepsilon)\gamma$ (green dotted line), for different infection parameter sets taken from Conway et al.\ \cite{conway13}. (D--F) Mean risk reduction in (A--C) for an antiviral acting to reduce $\gamma$ minus that for an antiviral acting to reduce $\beta$. Following Conway et al.\ \cite{conway13}, the risk of infection given that there is initially $n$ virions is $\text{risk}(n) = 1-[1-Q_c(1-\rhoV)]^n$ where $Q_c$ is the fraction of infectious virions in the exposure inoculum. The mean risk of infection is then $\sum_{n=0}^{N_\text{max}}\text{risk}(n)/(N_\text{max}+1)$ where $N_\text{max}$ is the maximum inoculum size determined such that the mean risk of infection without antivirals is $\sim$0.3\%. Finally, the mean risk reduction for a given efficacy $\varepsilon \neq 0$ is calculated as $ 100\% \cdot (\text{mean risk}_{\varepsilon=0}-\text{mean risk}_{\varepsilon \neq 0})/(\text{mean risk}_{\varepsilon=0})$. The infection parameters were taken from Conway et al.\ \cite{conway13}, i.e.\ $\tau_I = \unit{24}{h}$, $c = \unit{23/24}{h^{-1}}$, $n_I=1$, $p = pQ_a/24$, $R_0 = 8$, $\beta N_\text{cells} = (cR_0/\tau_I)/(p-R_0/\tau_I)$, $Q_c = 10^{-3}$, and our additional infection parameters were set to $\gamma = \unit{1}{cell/IV}$ and $\Vol = \unit{1}{mL}$.}
\label{conway13test}
\end{center}
\end{figure}

\end{appendix}
\end{document}